\documentclass{aa}
\usepackage{psfig}

\input epsf         % defines \epsfbox and supporting macros 
\begin{document} 
%\refereelayout

\title{A code for optically thick and hot photoionized 
media}

\subtitle{}

\author{Anne-Marie Dumont \inst{1},
        Arnaud Abrassart\inst{1}, 
        Suzy Collin\inst{1}}
\institute{
$^1$Observatoire de Paris, Section de Meudon, Place Janssen, F-92195 Meudon, 
France}

\date{Received  1999; accepted ...} %%  aa91.cmm

\thesaurus{ 03         % extragalactic astronomy
              (02.18.7;  % Radiative transfer,
               11.01.2;  % Galaxies:active,
               11.17.3;  % quasars:general
               13.21.1;  % Ultraviolet:galaxies,
               13.25.2)}  % X-rays:galaxies.

\offprints{Anne-Marie Dumont (Observatoire de Meudon)}

\maketitle{}

%\markboth{}

\begin{abstract}{We describe a code designed for hot media {(T $\ge$ }
a few 10$^4$ K), optically thick to Compton scattering. It computes the 
structure of a plane-parallel slab of gas in thermal and 
ionization equilibrium, illuminated on one or on both
sides by a given spectrum. Contrary  to the other photoionization codes, 
it solves the transfer of  the continuum and of the lines in a two stream 
approximation, without using the local escape probability formalism to 
approximate the line transfer. We stress the importance of taking into 
account the returning flux even for small
column densities (10$^{22}$ cm$^{-2}$), and we show that the escape
probability approximation can lead to strong errors in the thermal and 
ionization structure, as well as in the emitted spectrum, for a Thomson 
thickness larger than a few tenths. The transfer code is coupled with a
Monte Carlo code which allows to take  into account Compton and 
inverse Compton diffusions, and to compute the spectrum emitted up to 
MeV energies, in any geometry. 
Comparisons with {\sc{cloudy}} show that it gives  similar results for 
small column densities. Several applications are mentioned.}


\end{abstract}

\section {Introduction.}

A number of codes were built in the past to compute the structure and the
emission of photoionized media. With time, these codes became more 
and more sophisticated, and able to treat a larger number of situations. In 
the sixties they were specially designed for planetary nebulae and HII 
regions, i.e. for dilute and optically thin media (except in the Lyman 
continuum) ionized by the thermal radiation of a hot star, and were 
rapidly used also for ionization by a non thermal continua extending in
the X-ray range, i.e. for the Narrow Line Region of Active Galactic Nuclei 
(AGN). At the same time another generation of 
codes was developed for optically thin hot collisionally ionized 
media, and used for the solar corona, for supernova remnants, and
for the hot intergalactic gas. At the end of the seventies, observations of 
compact X-ray sources implying the presence of a Thomson thick medium 
incited Ross (1979) to develop a radiative transfer method for the 
continuum using a modified version of the 
Kompaneets equation. In the same paper he introduced the so-called 
``escape probability" approximation to take into account diffusion and 
absorption into the lines. This approximation was thereafter amply 
used for the Broad Line Region: this was the beginning of the
local escape probability era, with the fiducial paper of Kwan \&
Krolik (1981) aimed at studying the Broad Line Region of AGN 
(BLR),  immediately followed by a similar computation for the atmospheres of 
X-ray binary stars (Kallman and McCray 1982) from which {\sc{xstar}} derived. 
The most popular of these codes is {\sc{cloudy}}, designed by Ferland (cf. 
for instance Ferland \& Rees 1988) and continuously updated since this
time (cf. Ferland 1996, Ferland et al. 1998). These codes are not only
used for relatively ``cold" clouds like the BLR, but also for warm photoionized
media (Broad Absorption Line region in quasars and ``Warm Absorbers"
 in AGN, shock heated media, etc...).

At the other extreme (i.e. for dense, non irradiated, and semi-infinite 
media) model atmosphere codes have been developed since the
fifties with an emphasis put onto the computation of line transfer, 
which was treated completely, but with the LTE approximation. Non LTE 
{effects} were introduced in the sixties, and extensively studied in the 
seventies, with the state of the art being beautifully exposed in Mihalas book 
(1978).  Since this time many improvments have been made in
the numerical methods, in particular aiming at taking into account a large
number of atoms and levels, cf. for introductory reviews Rutten 1995, Hubeny
1997.

In the nineties appeared the urgent need for ``intermediate"
codes, valid for thick or semi-infinite dense media, eventually hot, irradiated
by a non thermal continuum extending in the hard X-rays, to cover the whole 
range of situations encompassed in AGN and in binary stars. Several 
codes were then built, mainly to compute the em-ission spectrum of accretion 
disks, irradiated or not by an X-ray continuum. For irradiated disks, they 
were either of ``photoionization 
type", using the Kompaneets equation (Ross \& Fabian, 1993, 
based on the Ross 1979 code), or coupling an existing photoionization code
to a Monte Carlo computation (Zycki et al. 1994), to take into account Compton 
diffusions. On the other hand, sophisticated model atmosphere codes were 
transformed to be applied to accretion disk structure and emission (Hubeny,
1990, Hubeny \& Hubeny 1997 and 1998), but without the external 
irradiation by an X-ray continuum.

Owing to the high optical thickness of the medium in several frequency 
ranges, such codes require that the
transfer of both the continuum and the lines be solved in an 
``exact" way, i.e. avoiding approximations such as local escape approximation 
(for the lines) or one stream approximation (for the continuum). Since 
the medium is generally dense and sometimes close to LTE,  they require 
that all processes and inverse processes be carefully
handled. Being irradiated by an X-ray continuum, the medium 
contains a large number of ionic species, from low to high ionization, 
which should all be introduced in the computation. Finally, the medium 
being hot and thick, not only
Thomson, but also Compton scattering, should be taken into account. 

 We have undertaken to build a code in order to satisfy these requirements. 
Precisely we have built several interconnected codes,
which allow more flexibility. The ensemble is far from being perfect 
and still contains several approximations which restrict its use, but 
we intend to improve it in the future.

We have presently four codes. One of them ({\sc{titan}}) is
designed to study the structure of a warm or hot thick photoionized gas, 
and to compute its emission - reflection - transmission spectrum from the 
infrared up to about 20 KeV. It solves the energy balance, the ionization 
and the statistical equilibria, the transfer equations, in a 
plane-parallel geometry, for the lines and continuum. 
Then, given the thermal and ionization stratification, the computation of 
the emitted spectrum from 1 KeV to a few hundre-ds KeV is performed  with 
{\sc{noar}}  which uses a Monte-Carlo method taking into
account direct and inverse Compton scattering (it allows also to
study various geometries). 
 {\sc{pegas}} adresses the case of more diluted and thin media, and is
similar to {\sc{cloudy}} with however some differences, and {\sc{iris}} is
specially devoted to the computation of the line fluxes, including very weak
lines, using the most recent available atomic data
and a full treatment of the statistical equilibrium equations for a 
great number of levels. These codes can be used to model a wide variety 
of astrophysical media, particularly those in which the energy transport is 
purely radiative, but also plasmas heated by other mechanisms 
(viscosity, shocks, energetic particules...). 
They have already been used in several published papers (Collin-Souffrin 
et al. 1996, Czerny \& Dumont 1998, Porquet et al. 1998, Abrassart 1999).

The three codes {\sc{noar}}, {\sc{pegas}} and {\sc{iris}}, are described 
in other papers (Abrassart 2000, Dumont \& Porquet 2000).
In this paper we present {\sc{titan}} and we discuss a few simple
cases, schematized by a slab of gas irradiated on one side or on both sides. 
As far as possible we will perf-orm comparisons with {\sc{cloudy}} 
in the range of parameters where both codes can be used, to assess the validity
of {\sc{titan}}. However detailed comparisons are not possible, as they use 
quite different transfer methods in particular. We will also show some 
examples of the capabilities of {\sc{titan}} in the parameter range which 
is not accessible to {\sc{cloudy}}.  

{\sc{titan}} is mainly designed to determine the structure (temperature and 
ionization state) and the continuum em-ission spectrum of a thick hot 
photoionized slab of gas. Owing to the representation of the ions made in the 
code it does not give an accurate determination of the weak lines. 
For a detailed determination of the line
spectrum of some ions, it should be complemented by {\sc{iris}}.

 We briefly summarize below the physical processes (Sect. 2),  the
transfer method and the iteration procedure (Sect. 3), the thermal 
equilibrium, (Sect. 4), focussing only on the aspects which are not 
treated in a standard way. The coupling of {\sc{titan}} and {\sc{noar}} is 
briefly described in Sect. 5, and some applications are presented in Sect. 6. 

\section{Physical processes}

In {\sc{titan}} the physical state of the gas (temperature, ion abundances 
and level populations of all ionic species) is computed at each depth, 
assuming stationary state, i.e. local balance between ionizations and
recombinations of ions, excitations and deexcitations, local energy balance
({equality} of heating due to absorption and cooling due to local 
emission), and finally total energy balance  (equality between inward and 
outward fluxes).

Due to the large range of density and temperature inside the medium, many
physical processes play a role at some place and should therefore be 
taken into account. Ionization equilibrium equations include radiative 
ionizations by continuum and line photons, collisional ionizations and 
recombinations, radiative and dielectronic recombinations, charge transfer 
by H and He atoms, the Auger effects, and  ionizations by high energy 
electrons arising from ionizations by
X-ray photons. Energy balance equations include free-free, 
free-bound and line cooling, and Compton heating/cooling.

The emission-absorption mechanisms for the continuum include free-free and 
free-bound processes, two-photon process, and Thomson scattering. 
Special care is given to recombinations to ground state, which are very 
important from an observational point of view in the X-ray range. They are 
treated differently according to the relative values of $kT$ to the 
photon energy bin, in order to get an accurate frequency dependence. 

Hydrogen and hydrogen-like ions are treated as 6-level atoms. 
Levels 2s and 2p are treated separately, while full l-mixing is assumed for 
higher levels. All processes including collisional and radiative ionizations 
and recombinations are taken into account for each level (cf. Mihalas 1978). 
Recombinations onto levels $n>5$ are not taken into
account, which amounts to assuming that the higher levels are in LTE
with the continuum, which is generally true in the conditions for 
which this code is presently used. In the future we plane to add several other
levels, and to sum the contributions of the higher levels as it is
done for instance in {\sc{cloudy}}. 
Level populations are then obtained as usual by matrix inversion.

In order to save computation time, non H-like ions are presently
treated with a rough approximation: interlocking
between excited levels is neglected and populations of the excited 
levels are computed separately using a two-level approximation. This 
approximation does not  predict correctly the details of the line spectrum,
since it neglects subordinate lines. Nevertheless recombinations
onto excited states are taken into account in the ionization equilibrium 
and the transfer of these photons is treated in an approximative way 
as proposed by Canfield \& Ricchiazzi (1980). We assume also that each 
recombination produces a resonant photon after cascades. 

The gas composition include 10 elements (H, He, C, N, O, Ne, Mg, Si, 
S, Fe), and all their ionic species are taken into account. 
Photoionization cross sections are fitted from Reilman \& 
Manson (1979) and Band L.M. et al. (1990), these values
being correct as far as neutral and once ionized
ions are not concerned, and it is the case in hot media. 
For total radiative and dielectronic recombination rates, we use Aldrovandi \& 
P\'equignot's data (1973). When possible, collisional excitation rates are 
taken from the Daresbury Report (1985). Most of data for iron come from 
Arnaud \& Raymond (1992), Kaastra \& Mewe (1993), and from
Fuhr et al. (1988). Ionizations by high energy electrons arising from 
ionizations by X-range photons are taken from Bergeron \& Souffrin (1973).
Inverse processes (except dielectronic recombinations) are computed through 
the equations of detailed balance. In the case of a gas close to LTE, we 
neglect dielectronic recombinations for consistency to insure the balance for 
each process. 
All induced processes are taken into account.

The equations are not recalled as they have been given in previously 
quoted papers.

\medskip

{\bf Iron K lines}

\medskip

These lines require special attention as they are intense in Seyfert nuclei, 
and they will be observed in detail in the future with Chandra and XMM. 
Though the iron K lines constitute a complex system
described in Band D.L. et al. (1990), we assume presently only one ``mean" 
line per ion with an oscillator strength equal to 0.4, as suggested by 
Band. More detailed computations are not required as far as the Doppler 
broadening of the lines (or Compton
broadening as well, see Abrassart 2000) is much larger than the distance 
between the lines, as it is the case in AGN. 

In other computations of the line fluxes it is generally assumed that 
resonant trapping of K$\alpha$ photons of FeXVII to Fe XXIII 
(Ross \& Fabian 1993) suppress completely these lines when the 
Thomson thickness of the emitting medium is larger than a given 
value, of the order of 0.02 (Zycki \& Czerny 1994). 
Here the transfer of these lines is handled in a standard way, with an 
additional term included in the statistical
and ionization equations to take into account the competition 
between radiative deexcitation and the Auger process. The population $N_i^k$ 
of the upper level of a transition K$\alpha$ of the ion $i$ is given by:

\begin{equation} 
N_i^kA_{ki}=y_{i-1}(N_{i-1}K_{i-1} + N_iB_{ik}J_{ik})
\label{eq-kalfa1} \end{equation} 

\noindent while the ionization equation for the same ion $i$ writes:

\begin{eqnarray} 
\label{eq-kalfa2}
%\ &\ &N_i[(P_i+K_i)+B_{ik}J_{ik}(1-y_{i-1})]
N_i[(P_i&+&K_i)+B_{ik}J_{ik}(1-y_{i-1})] \\
\nonumber
&=&N_{i+1}\alpha_i\ -\ N_{i-1}K_{i-1}(1-y_{i-1}). 
\end{eqnarray} 

$A_{ki}$ and $B_{ik}$ are the Einstein coefficients of the K$\alpha$ line, 
$J_{ik}$ is the mean intensity 
integrated over the line profile, $y_i$ is the fluorescent yield, 
$\alpha_i$ is the  recombination coefficient, and $P_i$ (respt. $K_i$)
is the photoionization rate of the ground state (respt. of 
the K-shell) of the ion $i$.
The second term on the left side of Eq. \ref{eq-kalfa2} is 
 generally only of the order of $10\%$ of the total photoionization rate. 

\section{Radiation transfer}

An important quantity will be used all along this paper, the illumination 
(or ``ionization") parameter. Among several definitions used in the 
literature, we adopt the following:

\begin{equation} 
 \xi_{\rm } = {4\pi F_{\rm inc} \over n_{\rm H}},
\label{eq-xi} 
\end{equation} 

\noindent where $F_{\rm inc}$ is the frequency integrated flux
 incident on one side of the slab and $n_{\rm H}$ is the hydrogen number 
 density at this surface. Note that it does not preclude the possibility 
of having  also a  flux incident on the back side of the slab. We call the 
attention on the fact that all along this paper we will {\bf mimic} a 
semi-isotropic and not a mono-directional illumination.

%\subsection{Previous methods}
%voir appendice

\subsection{Present method for the transfer of the continuum}

We want to study, on one hand media with an optical thickness 
larger than unity in a range of frequency rich in emission or in absorption 
lines, on the other hand  media with a very inhomogeneous structure. For 
instance if the illuminated side has a temperature of a few 10$^6$ K, while 
the temperature is  only a few 10$^4$ K on the back side of the cloud, the 
spectrum emitted by the illuminated side is mainly
formed in hot layers where the absorption coefficient is weak at all 
wavelengths and Thomson diffusion is dominant  (we shall call it the 
{\bf ``reflected spectrum" though it does not correspond to pure 
 reflection}). On the contrary the spectrum emitted by the backside 
of the slab is formed in cold layers 
where absorption must be carefully taken into account (we shall call it 
the ``outward spectrum"). Note also that, even if the incident 
illumination is mono-directional, the ``transmitted spectrum" is almost
isotropic since the Thomson thickness, $\tau_{\rm T}$, is larger than unity. 
It is therefore not differentiated from the 
``outward spectrum" even when the incident source is not located on the 
line of sight. It is possible to estimate an approximate transmitted 
spectrum using the expression $F_{\nu}^{\rm transm} = F_{\nu}^{\rm inc} 
exp(-\tau_{\nu})$, where $F_{\nu}^{\rm inc}$ is the incident flux,  
in an aim of comparison.

The transfer is treated with the Eddington two-stream approximation, 
i.e. the intensity is assumed to be constant in each hemisphere. 
The transfer equations can then be written :

\begin{eqnarray} {1 \over \sqrt{3}}{dI_{\nu}^{+} \over dz} 
&=& -(\kappa_{\nu} +{\sigma  \over 2})I_{\nu}^{+}+{\sigma  \over
2}I_{\nu}^{-}+\epsilon_{\nu} \\ 
\nonumber {-1 \over \sqrt{3}}{dI_{\nu}^{-} 
\over dz} &=& -(\kappa_{\nu} +{\sigma  \over
2})I_{\nu}^{-}+{\sigma  \over 2}I_{\nu}^{+}+\epsilon_{\nu} 
\label{eq-trans1}
\end{eqnarray}

\noindent where $z$ is the distance to the illuminated edge, $\kappa_{\nu}$ is 
the absorption coefficient, $\sigma$ is the diffusion coefficient - here 
it is due to Thomson scattering -  and $\epsilon_{\nu}$
is the {emissivity} (all these quantities are local and depend on the 
frequency $\nu$ except $\sigma$). Note that this approximation 
is closer to  the semi-isotropic case than to the normal case. 
We have already mentioned that it is
appropriate for the reflected and for the transmitted flux when the Thomson
thickness is larger than unity. It is also appropriate for a semi-isotropic
illumination, if the source of radiation is extended (like for instance in the
disk-corona model of AGN, or in the blob model of Collin-Souffrin et al. 1996).

The mean intensity $J_{\nu}$ is equal to $(I_{\nu}^{+}+I_{\nu}^{-}) / 2$,
while the flux F defined as usual by
$ \int I_{\nu} \cos{\theta}\ d\omega $ is equal to 
$(I_{\nu}^{+} - I_{\nu}^{-})\ 2\pi /\sqrt{3}$ ;
so that the reflected flux is {equal to} $I_{\nu}^{-} (0)\ 2\pi /\sqrt{3} $, 
and the outward flux to $I_{\nu}^{+} (H)\ 2\pi /\sqrt{3}$.
The optical depth and the total optical depth are defined  as :

\begin{equation} \tau _{\nu}(z)=\int_{0}^{z}
\sqrt{3}(\kappa_{\nu} +\sigma )dz'\ \ \ 
{\rm and}\ \  T_{\nu}=\tau_{\nu}(H) 
\label{eq-trans2} 
\end{equation}

\noindent where $H$ is the total thickness of the slab, and the source 
function as:

\begin{equation} S = {(\epsilon_{\nu} + \sigma J_{\nu})\over 
(\kappa_{\nu} + \sigma)} \label{eq-trans3} \end{equation}

%\noindent where $J_{\nu}$ is the mean intensity, $J_{\nu} = 
%(I_{\nu}^{+}+I_{\nu}^{-}) / 2$.

Both sides of the slab can be illuminated by an external radiation, so the
boundary conditions are, at $z=0$:

\begin{equation}  
I_{\nu}^+(0)={\sqrt{3}\over2\pi}\ F_{\nu}^{\rm inc},
\label{eq-trans3bis} \end{equation}

\noindent and at $z=H$:

\begin{equation}I_{\nu}^-(H)={\sqrt{3}\over2\pi}\ F_{\nu}^{\rm back}
\label{eq-trans3ter} \end{equation}

\noindent  where
$F_{\nu}^{\rm back}$ is the back-side flux (it is equal to 
zero in many of the following computations).

The formal solution of the transfer equations
% is :
%\begin{eqnarray} 
%\label{eq-trans4} 
%I_{\nu}^{+}(z) &=& I_{\nu}(0) e^{-\tau_{\nu}} +
%\int_{}^{}S_{\nu}(t) e^{-(\tau_{\nu}-t)}  dt \\
%\nonumber I_{\nu}^{-}(z) &=&
%\int_{}^{}S_{\nu}(t) e^{-(t-\tau_{\nu})}  dt 
%\end{eqnarray}
%\noindent and the variation of $I_{\nu}^{+}(z) $ 
between $z-\delta z$ and $z$ gives:

\begin{equation}I_{\nu}^{+}(z) = I_{\nu}^{+}(z-\delta z) e^{-
\delta\tau_{\nu}} + e^{-\tau_{\nu}}\int_{\tau_{\nu}-
\delta\tau_{\nu}}^{\tau_{\nu}}S_{\nu}(t) e^{+t} dt.
 \label{eq-trans5} \end{equation}

 Assuming that $S_{\nu}(t)$ is nearly constant in the interval $z,
z+\delta z$ , one gets: 

\begin{equation}I_{\nu}^{+}(z) = I_{\nu}^{+}(z-\delta z) 
e^{-\delta\tau_{\nu}} +
  {1-e^{-\delta\tau_{\nu}}\over 2}[S_{\nu}(z-\delta 
z)+S_{\nu}(z)]
 \label{eq-trans6} \end{equation}

\noindent with a similar equation for $I_{\nu}^{-}(z)$.

Guessing the initial value of $I_{\nu}^{-}(0)$ 
fails because calculations diverge, unless this initial value is 
provided with an extreme precision: the predicted value should differ 
from the real one at all frequencies by less
than $10^{-6}$. This problem has also been pointed out
by Coleman (1993). We therefore start to compute $I_{\nu}^{-}(z)$ 
from the back side with:

\begin{equation}
I_{\nu}^{-}(z) = I_{\nu}^{-}(z+\delta z) e^{-\delta \tau_{\nu}}+
{1-e^{-\delta\tau_{\nu}}\over 2}[S_{\nu}(z+\delta z)+S_{\nu}(z)]
 \label{eq-trans7} \end{equation}
o
Thus we calculate  $S_{\nu}(z)$ and the mean intensity $J_{\nu}(z)$ from
Eq. \ref{eq-trans3} and: 

\begin{eqnarray} 
J_{\nu}(z) &=& {I_{\nu}^+(z-\delta z) + I_{\nu}^-(z+\delta z) 
\over 2} e^{-\delta \tau_{\nu}} \\
\nonumber
&+& {(1-e^{-\delta\tau_{\nu}})\over 4} 
[S_{\nu}(z-\delta z) + 2S_{\nu}(z) + S_{\nu}(z+\delta z)]
\label{eq-trans10} \end{eqnarray}
 
\noindent assuming a constant $\delta z$ (actually it is not constant so the 
formulae are more complicated but do not deserve to be given here).

 The procedure is then the following. We divide the sl-ab into a set of 
plane-parallel layers with the density  $n_{\rm H}$ given in each layer.
Note that the code can easily be coupled with a prescription for the 
density or for the pressure, such as a constant pressure, or a pressure 
given by hydrostatic equilibrium. Presently the slab is divided in about 300 
layers. The geometrical thickness of each layer is sm-aller close to the 
surfaces than in the middle of the slab because of the rapid variation 
of the physical paramet-ers (the optical thickness, the temperature, 
the ionization state).

\medskip

\noindent - For each layer, starting from the illuminated side, 
where $I_{\nu}^+(0)$ is given by Eq. \ref{eq-trans3bis}, the ionization
and thermal balance equations are solved. The source function, the opacity
and the emissivity, are computed. As they depend on $J_{\nu}$, 
we use  Eq. \ref{eq-trans10} with the value of $I_{\nu}^+(z-\delta z)$ 
of the previous layer and the value of $I_{\nu}^-(z+\delta z)$ provided by
the previous iteration and given by Eq. \ref{eq-trans7} (see below).

\medskip

\noindent - $I_{\nu}^+(z)$ is transferred through the layer
according to Eq. \ref{eq-trans6}, while $S_{\nu}(z)$ and $\tau_{\nu}(z)$ are 
buffered for each layer and each frequency. 

\medskip

\noindent - when the back side of the cloud is reached, new values of 
$I_{\nu}^{-}(z)$  are calculated, starting from the back side 
where $I_{\nu}^{-}(H)$ is given by Eq. \ref{eq-trans3ter}, and using Eq. 
\ref{eq-trans7} with the previous values
of $S_{\nu}(z)$ and $\tau_{\nu}(z)$ ; actually we use the values given by
several previous iterations to accelerate the procedure.

\medskip 
  
\noindent - the whole calculation is repeated until convergence. It is
stopped when energy balance is achieved for the whole slab (i.e. 
when the flux entering on both sides of the slab is equal to the 
flux coming out from both sides).

\medskip
 
The drawback of this method is that it requires a large
number of iterations to get a complete convergence of the model. 
The iteration starts by assuming given values for $I_{\nu}^{-}(z)$. If 
this initialisation is not adequate, convergence problems are encountered. We
have chosen for the first iteration $I_{\nu}^{-}(z)$ equal to 
$I_{\nu}^{+}(z)$ multiplied
by a factor of the order of unity, which works reasonably well. 
Actually the convergence of $I_{\nu}^{-}(z)$ (as well 
as of the temperature and the ionization state) is reached within 
less than 1$\%$ after only a relatively small number of iterations 
in the layers corresponding to about one Thomson thickness. 
This property is interesting  as far as the outward spectrum is negligible, 
which is the case in  the EUV and X-ray range. It can also be used to 
determine the ionization state and the temperature of the layers where 
Compton diffusion takes place (cf. Sect. 5 about coupling with the
code {\sc{noar}}). 

\begin{figure}
\psfig{figure=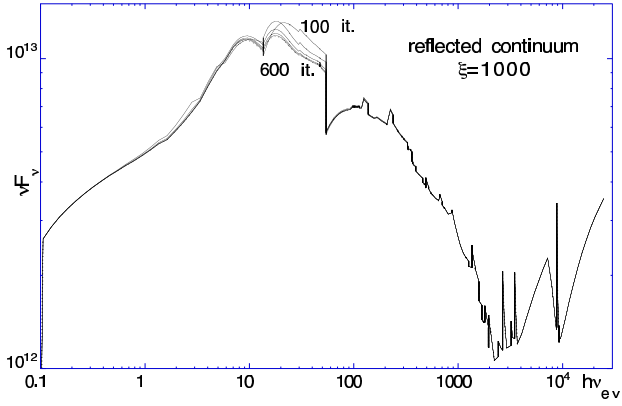,width=8.5cm}
\caption{The reflected continuum ${\nu}F_{\nu}$, in erg cm$^{-2}$ s$^{-1}$,
after a number of iterations varying from 100 to 600, displayed all 
100 iterations, for the reference model. We see that the spectrum does not 
vary at all after 400 iterations.}
\label{fig-Gstreflcont-it}
 \end{figure}
%\medskip

% To illustrate this point, 
Fig. \ref{fig-Gstreflcont-it} displays the
reflected continuum obtained after different numbers of iterations 
(for clarity we prefer to show here the continuum and not the complete 
spectrum).  
The model is a slab with the following characteristics:

\noindent - a constant density 10$^{12}$  cm$^{-3}$,

\noindent - a column density 10$^{26}$ cm$^{-2}$;

\noindent - it is illuminated on one side by an incident power law continuum 
proportional to $\nu^{-1}$ from 0.1 eV to 100 KeV,
 
\noindent - the illumination parameter is $\xi=10^3$ erg cm s$^{-1}$;
 
\noindent - there is no illumination on the  other side;
 
\noindent - the abundances are  (in
number): H: 1, He: 0.085, C: 3.3 10$^{-4}$,  N: 9.1 10$^{-5}$,
O: 6.6 10$^{-4}$, Ne: 8.3 10$^{-5}$, Mg: 2.6 10$^{-5}$, Si: 3.3 10$^{-5}$, S:
1.6 10$^{-5}$, Fe: 3.2 10$^{-5}$. 

In the following we will call this our 
``reference  model". All along the paper we will use the same density and 
the same spectral distribution of the illuminating radiation, so we will not 
specify these parameters anymore. 
Note that, for this peculiar spectral distribution, $\xi$ and the ratio
of incident ionizing photon number to hydrogen densities, U, are related by :
$ \xi = 8.21\ U\ \ln ({\nu_{\rm max}/\nu_{\rm min}})$  where $\nu_{\rm max}$ 
and $\nu_{\rm min}$ are the maximum and minimum photon frequencies.

We want to stress here the importance of the backward intensity. It cannot 
 be neglected as it participates to the ionization, to the
level population, and to the energy balance. For instance in a purely 
scattering medium the returning flux varies as  $F_{\rm inc} 
{\tau_{\rm T}/2 \over 1+\tau_{\rm T}/2}$, so when $\tau_{\rm T}$ is much 
larger than unity, all the incident flux is reflected. Thus the temperature of 
the first layer at the illuminated
side (which depends on the sum of incident plus returning radiation) 
should increase with the column density for a given incident 
flux (except possibly in cases of very high values of the 
illumination parameter).  

\begin{figure}
\psfig{figure=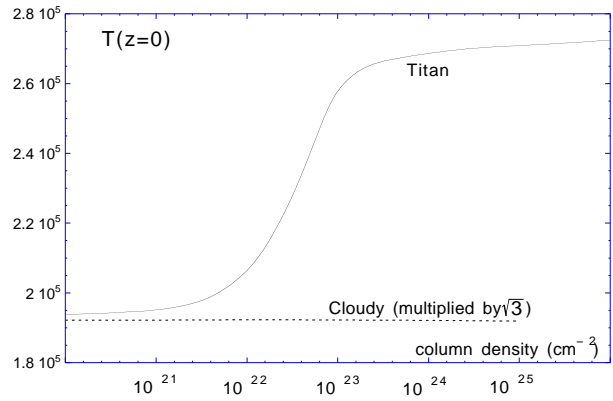,width=8.5cm}
\caption{Variation of the surface temperature, in K, with the column density, 
for a slab illuminated on one side by a power law continuum with $\xi=100$
erg cm s$^{-1}$. The surface 
temperature computed by {\sc{cloudy}} for $\xi$ multiplied by $\sqrt{3}$ 
is shown for comparison.}
\label{fig-GTitanCloudyT-col}
 \end{figure}

 Fig. \ref{fig-GTitanCloudyT-col} shows the dependence of the
surface temperature on the column 
density $N$. The model is similar to the reference one except that 
the illumination parameter is equal to 100 erg cm s$^{-1}$  and 
the  column density is varying. We see that the surface temperature is
constant up  to about $N=10^{22}$ cm$^{-2}$, but after it increases from 
2 10$^5$ K to 2.7 10$^5$ K for $N$ 
increasing from $10^{22}$ to $10^{24}$ cm$^{-2}$.
It saturates when a large fraction of the slab becomes close to LTE, and 
the returning flux becomes constant. With {\sc{cloudy}} (version 9004)  
the surface temperature does not depend on
the column density (for a correct comparison the illumination parameter in 
{\sc{cloudy}} must be multiplied by a factor $\sqrt{3}$ to account for the 
semi-isotropic approximation used in {\sc{titan}}).  

\begin{figure}
\psfig{figure=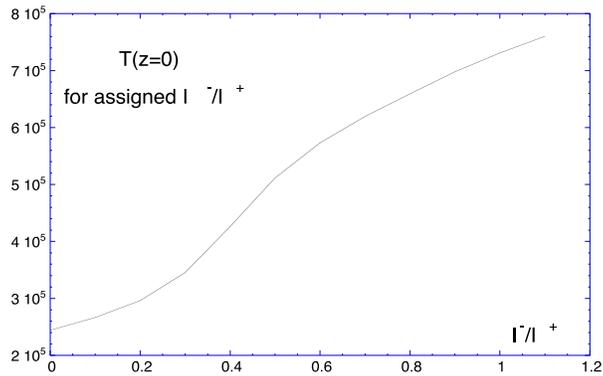,width=8.cm}
\caption{Dependence of the surface temperature, in K, on the ratio 
$I^-(0)/I^+(0)$ imposed a priori.}
 \label{fig-GT-fii}
\end{figure}

 Fig. \ref{fig-GT-fii} displays the dependence of the surface 
temperature on an assigned ratio $I^-(0)/I^+(0)$ independent of the 
frequency. We see that a modest returning flux (20$\%$) is sufficient 
to account for the  60$\%$ increase of the surface temperature  
reached for $N=10^{24}$ cm$^{-2}$,
with respect to a very small column density.  To understand this behaviour, 
one should remember that the spectral distribution of the reflected flux 
contains a much larger proportion of EUV radiation than the incident 
continuum, in particular in the lines. So the illumination parameter 
is increased by a larger amount than the ratio of the integrated
reflected over incident continuum. We can  also  guess that the  
increase of the temperature at the  illuminated  surface with respect 
to {\sc{cloudy}} will be compensated by a decrease at the back side, 
because of energy conservation (cf. Fig. \ref{fig-GtempTitClouc22-24}). 

\subsection{Present method for line transfer}

Radiative transfer of the lines is treated in the same way as the 
continuum, the line profile being represented by several frequencies 
distributed symmetrically around the line center.

We assume that even if large macroscopic velocities are present, they do 
not play any role in the line 
transfer, in other words that the thickness of the slab is much 
smaller than the scale length of the velocity gradient. We also 
assume that the lines do not overlap, which is valid as far as multiplets are 
treated globally, and if the microscopic 
turbulent velocity is not much larger than the thermal velocity 
(doing this we neglect line fluorescence due to the overlap of two 
lines, like the HeII Balmer$\beta$ line and the hydrogen Ly$\alpha$ line, 
whose wavelengths differ only by a fraction 4 10$^{-4}$). 

   Finally we assume presently complete redistribution in all the lines. 
We intend to include partial redistribution for
resonance lines like  Ly$\alpha$ in the near future. For the moment 
partial redistribution in these lines can be mimicked by assuming 
complete redistribution in a pure Doppler profile, as the line 
photons diffuse more in space than in frequency, owing to the 
absence of enlargment of the lower level. A Doppler profile can also 
be used for strongly interlocked lines, and in any case, it does not 
{differ} strongly from a Voigt profile when $\tau_0$ is smaller than 
$1/a$ so the medium is optically thin in the damping wings. 
Note that partial redistribution is less important in dense media, like
irradiated accretion disks in AGN, than in dilute media, owing to collisional
redistribution. 

%figure                                                    4
\begin{figure}
\psfig{figure=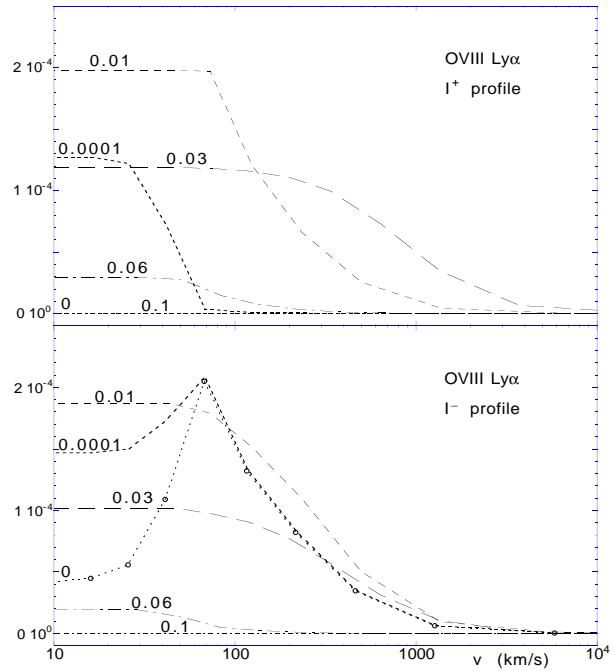,width=8.5cm,height=9cm}
\caption{Profile of $I_{\nu}^{-}$ and 
$I_{\nu}^{+}$ for OVIII Ly${\alpha}$, for the 
reference model. The curves are labelled by the value of $z/H$.}
\label{fig-GprofO8Lya}
\end{figure}

The line profile $\phi_{\nu}$ is a symmetrical Voigt function. 
The  Doppler width $\Delta\nu_{\rm d}$ and the damping constant depend on 
the temperature and on the density, and consequently vary with $z$. 
It may happen that the temperature decreases by two orders of 
 magnitude from the illuminated to the back side, so $\Delta\nu_{\rm d}$ 
decreases by one order of magnitude. This must be taken into account to 
choose the points in the profile. 
To solve the statistical equilibrium equations one needs to know the 
total line intensity weighted by the profile 
$ \int J_{\nu} \phi(\nu) d\nu $, 
and to treat the thermal and ionization equilibria one simply integrates the 
line intensity over the frequencies $ \int J_{\nu} d\nu $. 
For intense lines the first integral is dominated by the Doppler core, 
and the second by the wings. 
In the case of a saturated line with strong wings, the 
set of frequencies chosen to describe the line must cover a range from a few  
tenths $\Delta\nu_{\rm d}$ to a few hundreds $\Delta\nu_{\rm d}$ and 
at least 10 points are necessary for a correct representation of the profile.

Fig. \ref{fig-GprofO8Lya} displays for the reference model the 
variation with $z/H$ of the OVIII Ly${\alpha}$ profile in both 
directions, $I_{\nu}^{-}$ and  $I_{\nu}^{+}$, $J_{\nu}$ being the 
half sum of these profiles. We see that the profile varies 
considerably from the surface to the deeper layers.

\medskip

Now we are able to show the difference between the transfer 
treatment and the escape probability approximation for the lines 
discussed in the Appendix. In this aim, we have compared the divergence flux 
$\rho$ ( Eq. \ref{eq-div-flux}) with the escape probability towards the 
illuminated side $P_e(\tau_0)$, and with an approximate expression proposed 
by Rees et al. (1989) to take into account continuum 
absorption in the statistical and in the energy balance equations : 

\begin{equation}
\beta_{\rm eff} ={\kappa _c\over \kappa _c +\kappa _l} + 
{\kappa _l\over \kappa _c +\kappa _l}
{[P_e(\tau_0)+P_e (T_0-\tau_0) ]\over 2}
\label{eq-esc-probbis} 
\end{equation}
  
\noindent where $\beta_{\rm eff}$ is the total effective escape probability,
$\kappa _l$ and $\kappa _c$ are the absorption coefficients 
respectively at the line center  and in the underlying continuum, 
$T_0$ is the total optical depth of the slab at the line center, and 
$P_e(\tau_0)$ (respt. $P_e (T_0-\tau_0)$) are the escape probabilities 
towards the illuminated   (respt. the back) side. 
 
\begin{figure}
%\begin{figure*}
%\epsfxsize=8.8cm \epsfbox{Gnrbtout.eps}                   5
\psfig{figure=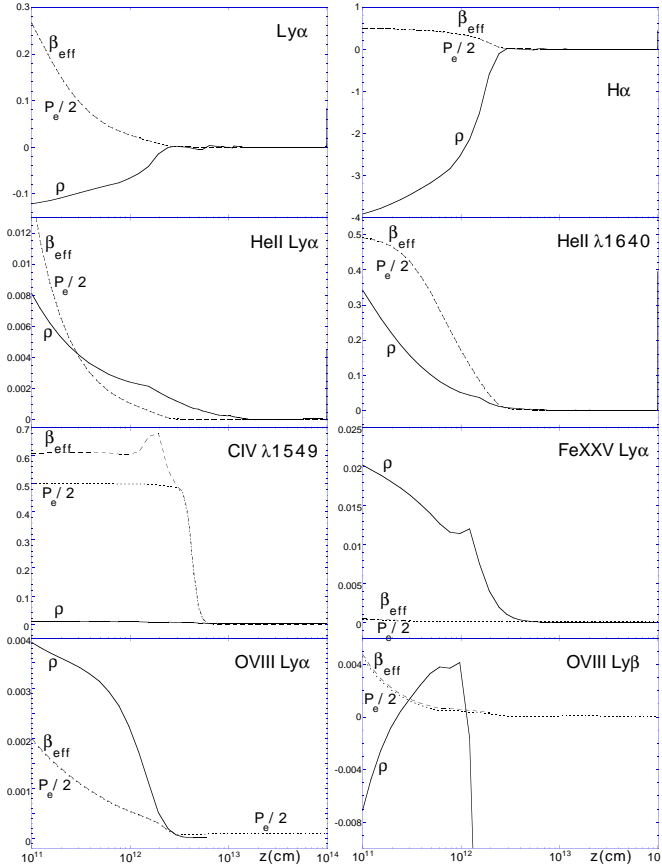,width=9cm}
\caption{Variation of $\rho$, $P_e(\tau_0) /2$ and $\beta_{eff}$ as 
functions of the distance to the illuminated side for several intense lines, 
for the reference model.}
\label{fig-Gnrbtout}
% \end{figure*}
 \end{figure}
  
$P_e(\tau_0)/2$ and $\beta_{\rm eff}$ have been computed with {\sc{titan}} and 
compared  to $\rho$ in Fig. \ref{fig-Gnrbtout}, as functions of the 
distance $z$ to the illuminated surface, for the reference model.

We see on this figure that $\beta_{\rm eff}$ and $P_e(\tau_0)/2$ 
are almost always identical. It is first due to the fact that the slab is very 
thick, so $P_e(\tau_0)$ is much larger than $ P_e(T_0-\tau_0)$ except 
very close to the back side, and second because $\kappa_c$ is always $\ll 
\kappa_l$, as all  the lines displayed on the figure are intense. But 
$\beta_{\rm eff}$ differs considerably from 
$\rho$ in the region where the lines are formed (i.e. the region 
where $\rho$ is not equal to zero and where the ion is present). In particular 
$\rho$ is negative in an important fraction of the line formation region for 
H$\alpha$,  L$\alpha$, and OVIII L$\alpha$, while $\beta_{\rm eff}$ is a 
positive quantity. 

\begin{table}
\psfig{figure=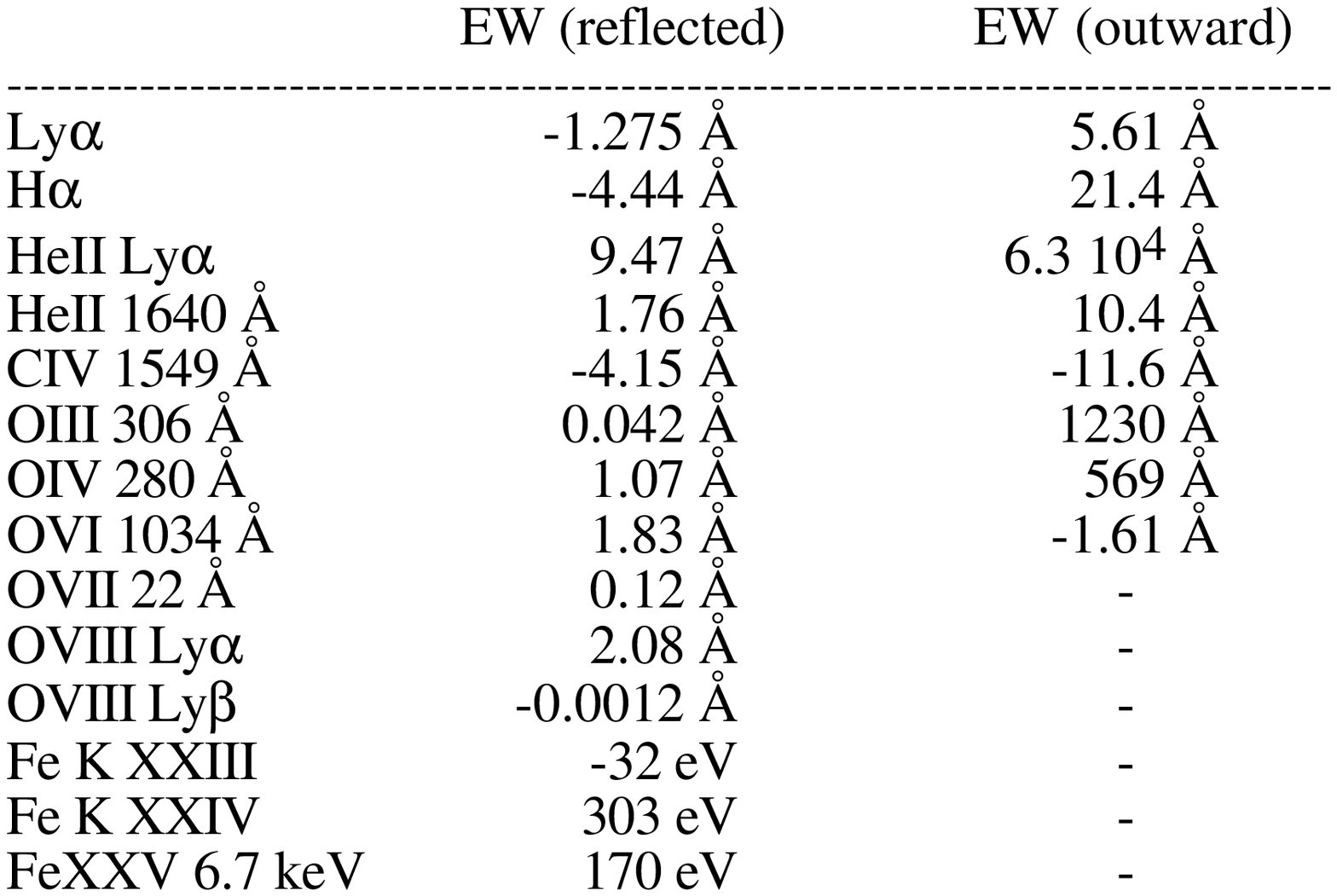,width=8.5cm,height=5cm}
\caption{ Equivalent widths of a few intense lines (negative for
absorption, positive for emission) in the reflected and 
in the outward spectra, for the reference model.}
\label{table-1}
 \end{table}

As a consequence the line spectrum is not correctly computed when 
escape probabilities are used. This can be seen on Table \ref{table-1} 
which gives the equivalent widths (negative if the line is in absorption,
positive if it is in emission) of some intense lines from both  sides 
of the slab. Several of these lines are in absorption, either from the 
illuminated or from the dark side, which is impossible with the 
escape probability approximation. 
%Note that the total optical thickness 
%of the continuum at the frequency of the lines is always 
%very large, so clearly Eq. \ref{eq-F} (which is used in the escape 
%probability approximation) is not valid.

\bigskip
Presently the radiative transfer is solved for 300 frequencies in the 
continuum from 0.01 eV to 25 keV, including all the ionization edges, 
and for 440 spectral lines including the fluorescence
Fe K lines, the integrals over the line profiles being achieved
using 15-point Gauss-Legendre quadrature, together with a change of 
variable.

%figure                                                           6
\begin{figure}
\psfig{figure=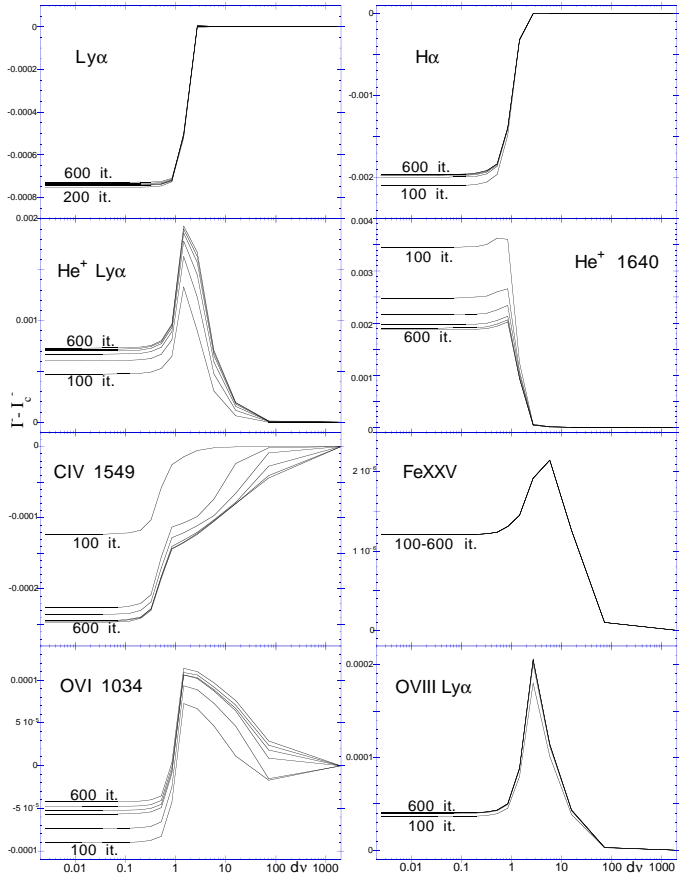,width=9 cm}
\caption{Profiles of some intense lines emitted by the bright side after 
a number of iterations varying from 100 to 600, displayed all 100 iterations, 
for the reference model.}
\label{fig-stprofraies-it}
 \end{figure}

\bigskip
{\bf Remarks on the convergence of the method}
\medskip

It is well-known that the lambda iteration method used here 
(actually mainly for historical reasons, because we st-arted to build 
a nebular code with the line escape probability approximation)  
converges extremely slowly for intense lines in stellar
atmospheres, and we intend to replace it by the Accelerated Lambda  
Iteration method as soon as possible. The convergence of $J_{\nu}$ 
is achieved  after 600 iterations, for all lines except
a few ones and for all continua. This can be checked on 
Fig.  \ref{fig-stprofraies-it} which displays for the same lines as 
Fig.  \ref{fig-Gnrbtout} the profiles of the lines emitted by 
the bright side, after different number of iterations, for the 
reference model. As a consequence, the structure of the slab (T, ionization)  
does not change at all after a few hundreds iterations. 
Indeed our case is more  favorable than stellar atmospheres.  The
explanation is double: 1. the relatively small density of the medium 
compared to a stellar atmosphere, which decreases comparatively the 
influence of collisional excitations, and 2. the fact that in a 
photoionized plasma the temperature is low for a high ionization degree, 
so the excited levels lie at very high energies compared to the
thermal energy. Consequently the populations of these highly ionized ions 
are  smaller than at Boltzman equilibrium  with the ground level, and the 
lack of convergence for excited level populations does not have a strong 
influence on the ionization state of the corresponding ion. We have also 
checked that the  same structure is obtained for different initial
boundary conditions. All this implies that the continuum emitted from 
both sides (including for instance the Lyman discontinuity) is 
completely converged, and only the few non converged lines are not 
well computed, in particular those emitted by the back
side (but they are negligible in the  whole energy balance). Since we 
do not claim to compute a correct line spectrum  before the subordinate 
lines are implemented in the code, we think that this  approximation is 
sufficient for our present purpose.

\section{Energy balance}

{\sc{titan}} has been designed for stationary radiatively heated media, so the 
energy conservation implies the balance of radiative flux. However it 
can be used also for other cases out of radiative equilibrium, 
for instance by imposing the temperature of the medium, or by adding 
a non radiative energy in each layer, such as the viscous flux of an accretion 
disk. We describe here how the code works in a  purely radiative case.

We first stress the fact that the energy balance 
should be realized both {\it locally} and {\it globally}.

 The local energy balance equation writes:

 \begin{equation} 
\int_{}^{} {dF_{\nu}\over dz} d\nu\ =\ 0\ =n_en_H[\ -
\Gamma_{\rm 
tot}\ +\ 
\Lambda_{\rm tot}],
 \label{eq-energy1} \end{equation}

\noindent where $n_en_H\Gamma_{\rm tot}$ and $n_en_H\Lambda_{\rm tot}$ are 
the total heating and cooling rates per unit volume.
The temperature of a layer is computed through an iteration process 
until $\Gamma_{\rm tot}$ and $\Lambda_{\rm tot}$ are equal. 

%figure                                                     7
\begin{figure}
%\begin{figure*}
\psfig{figure=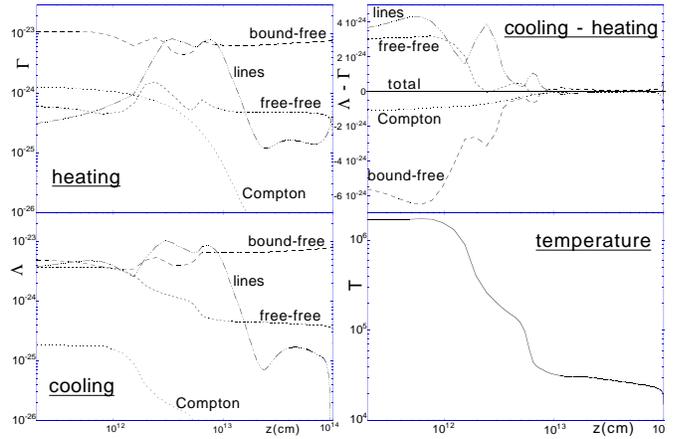,width=9cm}
\caption{Dependence of the local cooling, heating, and net 
cooling-heating rates, in erg cm$^{3}$ s$^{-1}$, and of the temperature, 
in K, on the distance to the illuminated side $z$, for the reference model.} 
\label{fig-Gst-local-balance-T}
%\end{figure*}
\end{figure}

As an example, Fig. \ref{fig-Gst-local-balance-T}
gives the dependence of the cooling and heating 
rates, and of the equilibrium temperature, on the distance from the
illuminated side $z$, for the reference model. We note that in this 
model Compton heating-cooling is
not very important, and that bound-free contributes as a
heating term, compensated by a cooling term due to lines and free-free 
processes. Note also that a quasi LTE is reached at a few Thomson depths in 
this model (it is reached at a smaller
optical depth for a sm-aller value of $\xi$). Above this value of $z$ the 
temperature is almost constant, except close to the back side where it 
decreases more rapidly due to the leakage of photons.  

%figure                                                    8
\begin{figure*}
\psfig{figure=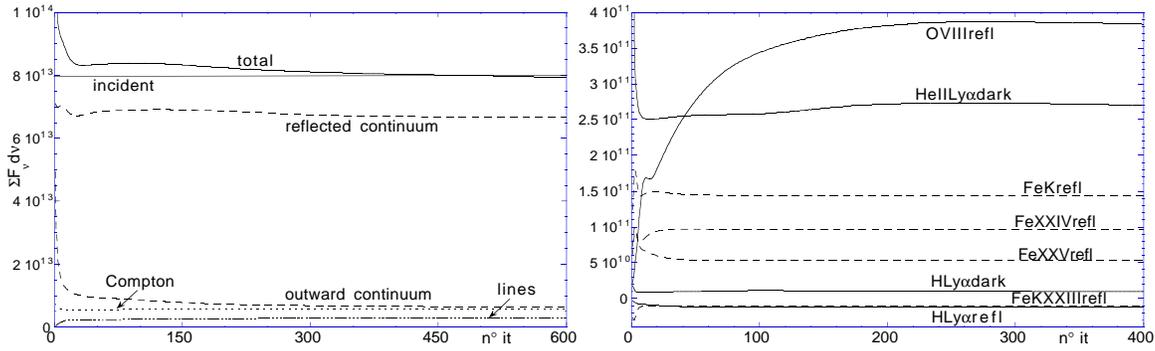,width=16.0cm}
\caption{Dependence of the total line and continuum fluxes and 
of the fluxes of a few intense lines on the number of iterations, 
for the reference model.} 
\label{fig-Gst-global-balance-it}
\end{figure*}

  The global energy balance is reached through an iteration procedure,
when the total flux emitted by both sides of the slab, 
$F_{\rm tot}^{\rm em}$, is equal to the total flux incident
on both sides, $F_{\rm tot}^{\rm inc}$. Actually this is the most 
important convergence criterium and it requires many iterations. 
To illustrate the problem, Fig. \ref{fig-Gst-global-balance-it} 
displays the flux of a few intense emission lines and of the 
total flux in the lines and continuum, for the
reference model, as functions of the number of iterations. 
We see  that after about 200 iterations the global energy balance is
reached within about one percent.

%figure                                                    9
\begin{figure*}
\psfig{figure=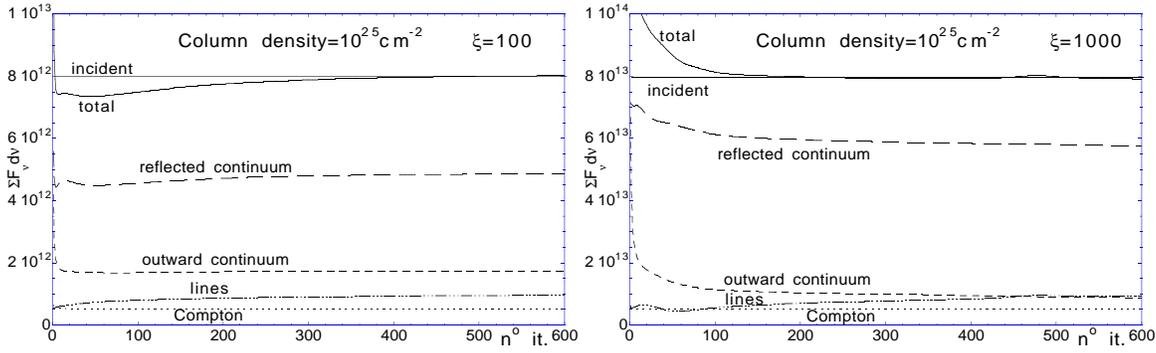,width=16.0cm}
\caption{Dependence of the total line and continuum fluxes on the 
number of iterations, for  a column density of 10$^{25}$ cm$^{-2}$, and 
$\xi=10^2$ erg cm s$^{-1}$ and $\xi=10^3$ erg cm   s$^{-1}$. } 
\label{fig-Gbilc25x2-3}
\end{figure*}

In the same way, Fig.\ref{fig-Gbilc25x2-3} shows that the energy 
balance with {\sc{titan}} is reached within one percent
after 200 iterations for two models with a column density equal to 
10$^{25}$ cm$^{-2}$, and with an illumination parameter equal to 
$\xi=100$ and $\xi=1000$ erg cm s$^{-1}$. {\sc{cloudy}} or {\sc{xstar}}
have been used several times in this range of parameters 
(Martocchia \& Matt, 1996,  Zycki et al. 1994, for instance), 
but it is not sure that the global energy balance is achieved
owing to the escape treatment of the lines, as discussed above.

%\medskip

Finally we must recall that the neglect of subordinate lines (except
in hydrogen-like ions) could  lead to a substantial increase of the 
energy losses and consequently to a change of the thermal balance, 
as these  lines have a smaller optical thickness than resonant  lines, and 
therefore can escape more easily from the deep layers. Since in a 
photoionized plasma the temperature is relatively low and  the excited level 
populations of highly ionized ions are generally  small, one could expect 
only a small influence on the overall spectrum. 

%figure                                                    10
\begin{figure}
\psfig{figure=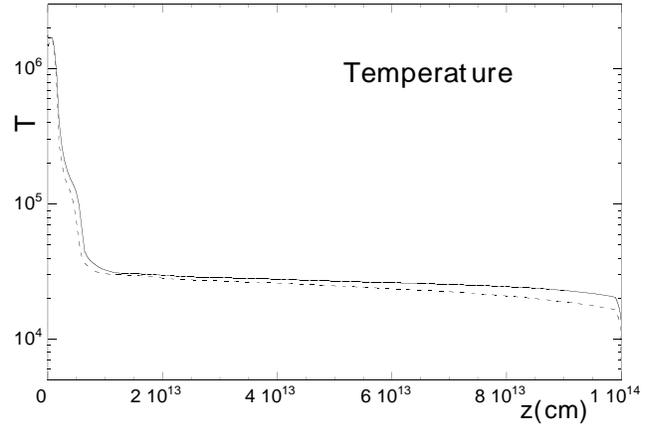,width=8.5cm}
\caption{Temperature in K versus $z$ with (full 
line) and without (dashed lines) full five-level interlocked 
hydrogen-like ions,  for the reference model.} 
\label{fig-Gst-hyd-T}
\end{figure}

\begin{figure*}
\psfig{figure=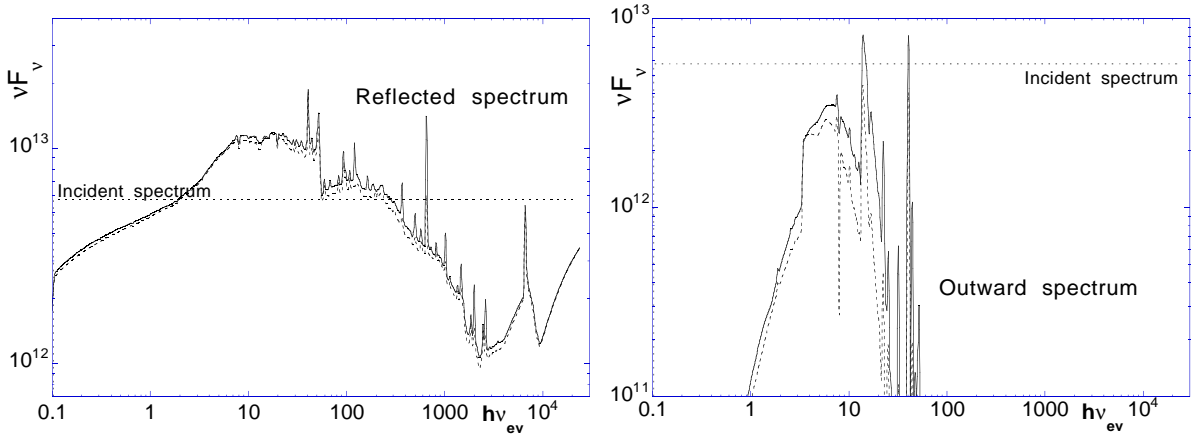,width=16.0cm}
\caption{Reflected and outward spectra, with (full 
line), and without (dashed lines) full five-level interlocked 
hydrogen-like ions, for the reference model. The spectra ${\nu}F_{\nu}$ are 
displayed with a line resolution of 30, in erg cm$^{-2}$ s$^{-1}$.} 
\label{fig-Gst-hyd-spectres}
\end{figure*}

 To check this point we compare  in Figs.  \ref{fig-Gst-hyd-T}
and \ref{fig-Gst-hyd-spectres} the temperature and
the emitted spectrum for the reference model, with and without
taking into account interlocking between excited levels in hydrogen 
like ions (note that interlocking is included for hydrogen and helium 
in both cases). These ions are the most abundant ones in the 
region emitting the reflected spectrum, in our reference model.  Fig. 
\ref{fig-Gst-hyd-T} shows that the temperature is slightly shifted in layers 
corresponding to a Thomson thickness of a few. As a consequence 
the emitted spectrum is also different, but not by an important 
amount. We can infer a similar behaviour for 
smaller illumination parameters, even if the abundant {ions} are less
ionized, as the temperature is also lower. For a higher illumination
parameter the lines contribute less to the  energy balance.  

%figure                                                    11

Since the detailed computed line spectrum of hydrogen-like ions is 
different with the two treatments, contrary to the overall spectrum, 
it is mandatory to treat correctly the excited levels for all the 
ions in order to get a correct line spectrum. Note
that in this respect {\sc{cloudy}} is presently doing much better than 
{\sc{titan}}, as it includes subordinate lines and  more levels for each ion.

\medskip

\noindent{\bf Compton heating and cooling}

\medskip

 The Compton heating-cooling term is equal to:

 \begin{equation} 
\Gamma-\Lambda={\sigma_T\over m_ec^2}{1\over 
n_H}4k(T_{\rm Comp}-T)\int_{}^{}{4\pi J_{\nu}d\nu}
 \label{eq-Compton} \end{equation}

\noindent where $T_{\rm Comp}$ is the Compton temperature. We give here this 
well-known equation only to recall that it depends (as well as  
$T_{\rm Comp}$) on the spectral distribution of the {\it local} flux, so it 
requires a careful computation of this flux (including the lines). 

Below 25 KeV, $J_{\nu}$ is provided directly by {\sc{titan}} in each layer.
Above 25 KeV, $J_{\nu}$ is not calculated by {\sc{titan}}, but by the
code {\sc{noar}}, as the energy shift
of the photons due to Compton scattering is no more negligible. In 
this case the results from {\sc{noar}} are also used to compute the exact
value of the Compton heating-cooling term. This is discussed in 
more detail in the next section.

\section{Coupling with the Monte Carlo code}

In order to compute the effect of Compton scattering on the high 
energy part of the reprocessed radiation, and particularly on the reflected 
spectrum above 1 keV, and to determine precisely the Compton 
heating-cooling rate entering in the energy balance equation, we need
another numerical approach. The transfer scheme in {\sc{titan}} is based 
on frequency arrays, and inelastic scattering would introduce a 
redistribution on energy bins making the code much heavier. In this aim 
Abrassart (2000)  developped a Monte Carlo code, {\sc{noar}}. The 
asset of such an approach is also that it enables to investigate 
an arbitrary geometry and to determine the angular dependance of 
the observed spectra. These aspects were  difficult to tackle with 
a code such as {\sc{titan}}. Moreover, it allows to easily extract time 
variability information.

 {\sc{noar}} is described in detail in Abrassart (2000). We 
 explain here only how it is coupled with {\sc{titan}}. 
It shares the same opacity  data and includes the same set of 102 
ions, for the 10 elements of highest cosmic abundances. Given all the 
ionisation degrees provided by {\sc{titan}}, it computes 
absorption cross sections in each layer. Free-free absorption 
is also taken into account. {\sc{noar}} includes the 
recombination continua of hydrogen and helium like ions, but does not 
include line emission, except fluorescence lines.
The proper yields to account for the competition with the 
Auger effect are used for these lines. Fluorescence of Fe XVII-XXIII 
is suppressed by resonnant trapping.
For the pseudo fluorescence of H and He-like species,  a simplifying 
prescription is adopted which includes only the two most probable outcomes 
of a K shell photoionization, i.e. direct recombination on ground level
or L-K transition.  The spectral distribution of the recombination continuum is
determined by the local temperature in the ``current" slab. 

{\sc{noar}} takes into account direct and inverse Compton scattering. The 
method used for modelling Compton scattering is basically the one 
described in Pozdniakov, Sobol \& Sunyaev (1983) and in Gorecki 
\& Wilczewski (1984), with a different use of statistical weight, 
because of the competition with photoelectric absorption.
When the high energy cut-off of the incident continuum is of the 
order of 100 KeV,
the reflected spectrum above 10 KeV exhibits a ``Compton  hump" peaking at 
about 30 KeV (Lightmann \& White 1988). This spectral region of optimal 
continuum albedo occurs where 
the sum of photoelectric absorption and Compton losses is minimum. 
This feature depends weakly on the ionization state, although its prominence 
in an observed X-ray spectra, and notably its level with respect to the 
iron line, is sensitive to $\xi$ (see Fig. \ref{fig-Gtitnoarx2-3-4}) 
as the soft X-ray albedo. Note that the shape and extension of the hump 
depends on the high energy cut-off of the incident spectrum, which 
is 100 KeV all along this paper. 

Fluorescent lines are significantly Compton broadened
for rather high illumination parameter, above $\xi\sim 10^3$ erg cm s$^{-1}$. 
The broadening is asymmetyric, the profile is skewed toward the red. At still 
higher $\xi$ and temperature, a blue wing appears, due to 
the inverse Compton effe-ct. The extend and importance of both wings 
depends on the optical depth of the scattering medium. The red wing 
depends only on the frequency of the line (i.e. on the ionization state of
iron) whereas the blue  one only depends on the 
temperature of the scattering medium.

{\sc{titan}} and {\sc{noar}} lead to almost identical reflected spectra in 
the 1-20 KeV range (they differ only by Compton broadening of lines and 
of photoelectric edges), and {\sc{noar}} allows to obtain the spectrum in 
the higher energy range. This spectrum is used for the global energy balance.

Another use of {\sc{noar}} is to provide {\sc{titan}} with the local Compton 
gains and losses in each layer. This is necessary, because
Compton heating-cooling is dominated by energy losses of photons $>$ 25 Kev,
in particular for high values of $\xi$ when Compton heating-cooling 
is dominating the energy balance and should be computed in an exact way. 

%figure                                                   12
\begin{figure}
\psfig{figure=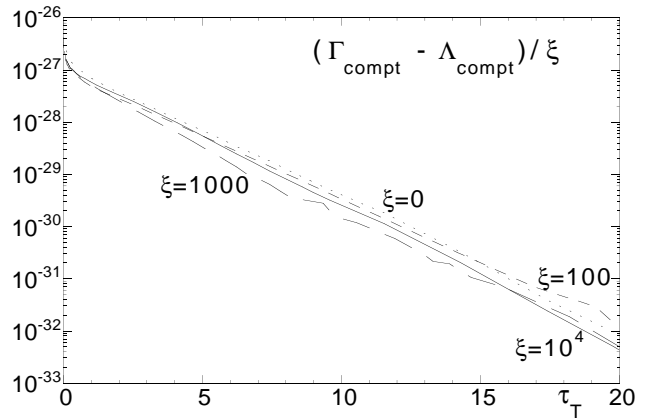,width=8.5cm}
\caption{Dependence of the ratio ${\Gamma_{\rm Comp}- 
\Lambda_{\rm Comp} \over\xi}$ on the Thomson
depth, for different values of $\xi$ in erg cm s$^{-1}$, the other 
parameters being the same as in the reference model.} 
\label{fig-Compton}
\end{figure}

%figure                                                   13
\begin{figure}
\psfig{figure=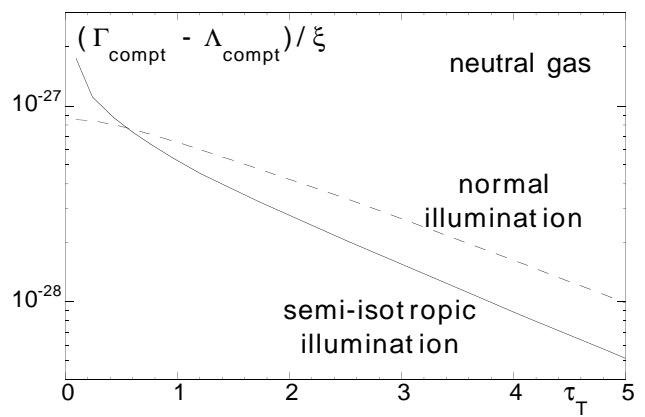,width=8.5cm}
\caption{Dependence of the ratio ${\Gamma_{\rm Comp}- 
\Lambda_{\rm Comp} \over\xi}$ on the Thomson
depth,  for a neutral gas and for an isotropic or perpendicular 
illumination. Except $\xi$ the parameters are the same as in the 
reference model.} 
\label{fig-Compton-neutre}
\end{figure}

As an example Fig. \ref{fig-Compton} gives the dependence of the 
ratio $(\Gamma_{\rm Comp}- \Lambda_{\rm Comp})/\xi$ on the Thomson
depth, for different values of $\xi$, the other parameters being the 
same as in the reference model. We see that the ratio does not depend on 
the illumination parameter if $\xi \le 10^4$ erg cm s$^{-1}$.  This is 
expected since the Compton heating-cooling is dominated by high energy 
photons, which do not care about the ionization state. It also means that 
inverse Compton is negligible as a cooling process (though it has an 
influence on the line spectrum), otherwise ${\Gamma- 
\Lambda \over\xi}$ would depend on the gas temperature and 
therefore on $\xi$. This is due to the relatively low temperature of 
the gas compared to the mean energy of the photons. This property 
breaks for very high illumination parameters, since low energy
photons are then able to penetrate in the deepest layers and to play 
a role in the cooling 
rate, and the temperature of the gas is close to $T_{\rm Comp}$.

The deposition of energy shown on Fig. \ref{fig-Compton} corresp-onds 
to a semi-isotropic distribution of the incident 
radiation. It is different in the case of a 
mono-directional radiation. This is shown on Fig. 
\ref{fig-Compton-neutre} which displays $(\Gamma_{\rm Comp}- 
\Lambda_{\rm Comp})/\xi$ for a neutral gas in the case of an isotropic 
and of a perpendicular illumination.  When the radiation field is 
semi-istropic there is a rapid decrease of ${\Gamma- \Lambda 
\over\xi}$ for small values of $\tau_{\rm T}$, due to the small distance 
covered by the photons. This is compensated by a smaller (by about 
a factor two) deposition of energy in the deeper layers. 

The Compton heating-cooling rate obtained with {\sc{noar}} is then fitted 
analytically as a function of $z$, which is transferred to {\sc{titan}}.

\section{Some applications}

A full grid of models will be published elsewhere (Dumont \& Abrassart 2000). 
Here we discuss only a few interesting cases. 

\subsection{Optically thin hot medium: the Warm Absorber in AGN}

The Warm Absorber (WA) is a hot medium located on the line of sight of 
the X-ray source in AGN. It is responsible of the absorption edges of 
OVII and OVIII observed in their soft X-ray spectrum in about 50$\%$ 
of them. It might also constitute the ``mirror" invoked in the Unified 
Scheme to account for the polarized Broad Lines observed in Seyfert 2 
galaxies (Antonnucci \& Miller 1985). It has been modelled in many papers, 
using {\sc{ion}} (Netzer 1993, 1996), {\sc{xstar}} (Krolik \& Kriss 1995), 
{\sc{cloudy}} (Nicastro, Fiore \& Matt 1999 for the most recent paper), 
{\sc{pegas}} and {\sc{iris}} (Porquet et al. 1999). 
This medium  has a column density of 10$^{21}$ to 10$^{24}$ cm$^{-2}$, 
and a typical illumination parameter of $\xi=100$ erg cm s$^{-1}$. 
Porquet et al. have shown that its density should be at least 
10$^{10}$ cm$^{-3}$ to avoid the emission of too strong coronal lines. 

As the WA is relatively thin ($\tau_{\nu}\ll 1$ for the 
continuum in a large range of frequencies), it is possible to use 
the line escape probability formalism. This is however limited to the case 
where $\tau_{\nu}$ is smaller than unity in the continuum underlying 
intense lines. This specific problem will 
be discussed in detail in the paper describing the code {\sc{pegas}} 
(Dumont \& Porquet 2000). 
A correct computation of the backward 
flux is also required. We have shown indeed that  the 
temperature depends on the intensity of this backward flux. 

We have run several models corresponding to the conditions 
of the WA, i.e. for $\xi=100$ erg cm s$^{-1}$ and 
column densities varying from 10$^{22}$ to 10$^{24}$ cm$^{-2}$ 
(the other parameters being the same as 
in the reference model). We have compared the results to those 
obtained with {\sc{cloudy}} for the same models, to check that they 
are similar at low column densities, and to determine at 
which column densities they begin to differ strongly. 
Note that a detailed comparison between {\sc{titan}} and {\sc{cloudy}} is not 
possible, as these codes differ both in the transfer method and in 
the atomic data (for instance owing to the semi-isotropic 
illumination,  a given column density corresponds to a 
larger optical thickness with {\sc{titan}} than with {\sc{cloudy}}), so only 
trends can be obtained. 
   
%figure                                               14
\begin{figure}
\psfig{figure=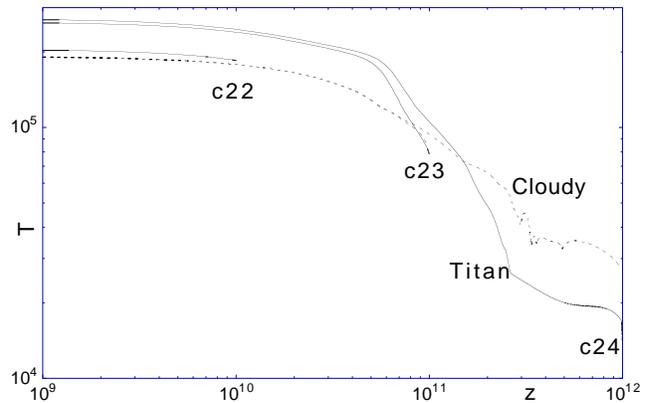,width=8.5cm}
\caption{Dependence of the temperature in K on the distance to the 
illuminated side, for $\xi=100$ erg cm s$^{-1}$ and 
column densities varying from 10$^{22}$ to 10$^{24}$ cm$^{-2}$. 
The temperature obtained with {\sc{cloudy}} is shown for comparison.}
\label{fig-GtempTitClouc22-24}
 \end{figure}

%figure                                               15
\begin{figure}
\psfig{figure=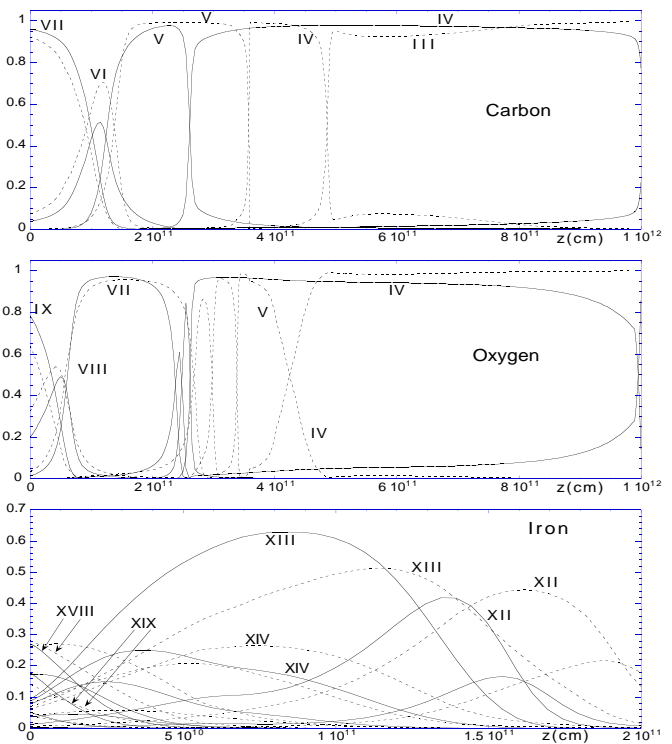,width=9cm,height=10cm}
\caption{Fractional abundances of Carbon, Oxygen, and Iron, 
computed by {\sc{titan}} (full lines) and {\sc{cloudy}} (dashed lines),
as functions of the distance to the illuminated side, for
$\xi=100$ erg cm s$^{-1}$ and a column density equal to 10$^{24}$ cm$^{-2}$. 
Note that the absissa scale for Iron differs from the others.}
\label{fig-ionisation-24}
 \end{figure}

Fig. \ref{fig-GtempTitClouc22-24} displays the dependence of 
the temperature on $z$, obtained with {\sc{titan}} and with {\sc{cloudy}}. As 
discussed previously, it is higher with {\sc{titan}} at the 
illuminated side for a high column density,  owing to the returning 
flux, and it is lower at the dark side. 

\begin{figure}
\psfig{figure=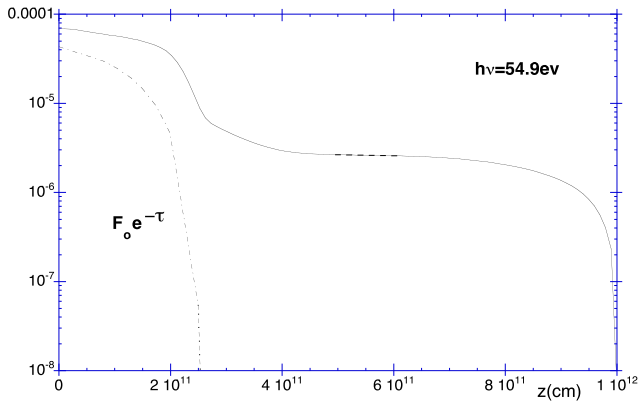,width=8.5cm}
\caption{Variation of the flux at 54.9 eV (the OIII-OIV ionization edge) 
as a function of the depth, and comparison with the transmitted flux.}
\label{fig-comparaison-cont}
 \end{figure}

The ionization state (of fundamental importance 
for the study of the WA) and the emission spectrum, are also 
different with {\sc{titan}} and with {\sc{cloudy}} at high column densities. 
Fig. \ref{fig-ionisation-24} shows the ionization state of Carbon, 
Oxygen and Iron for a column density of 10$^{24}$ cm$^{-2}$, computed with 
{\sc{titan}} and with {\sc{cloudy}}. While the ionization state close to 
the illuminated side is quite similar (in particular if one remembers that 
there is a factor $\sqrt{3}$ in the optical depth), it is very different 
in the deepest layers, where {\sc{cloudy}} leads to a smaller degree of 
ionization than {\sc{titan}} (CIII and OIII instead of CIV 
and OIV). This effect cannot be explained by the difference in optical 
thickness, which acts in the opposite direction. It is certainly due to 
the different treatments of the diffuse continuum, and perhaps of the lines. 
With {\sc{titan}} we have checked that the diffuse continuum at large 
depths is sufficient to ionize OIII into OIV. To illustrate this point Fig.
\ref{fig-comparaison-cont} displays the variation of the flux at 54.9 eV 
(the OIII-OIV ionization edge) as a function of the depth. It shows that 
the diffuse flux is still very intense and sufficient to ionize OIII at 
large depths (contrary to the transmitted flux which is very weak). 
Note that this result {\it depends strongly on the way the spectral 
distribution of the diffuse flux is computed}. If it would be reduced 
to a unique frequency at the recombination frequency of He$^{++}$, 
54.4 eV, (as it is generally the case with the ``On The Spot"  
approximation), there will be no photons able to ionize OIII.

%figure                                            16

%figure                                            17
\begin{figure*}
\psfig{figure=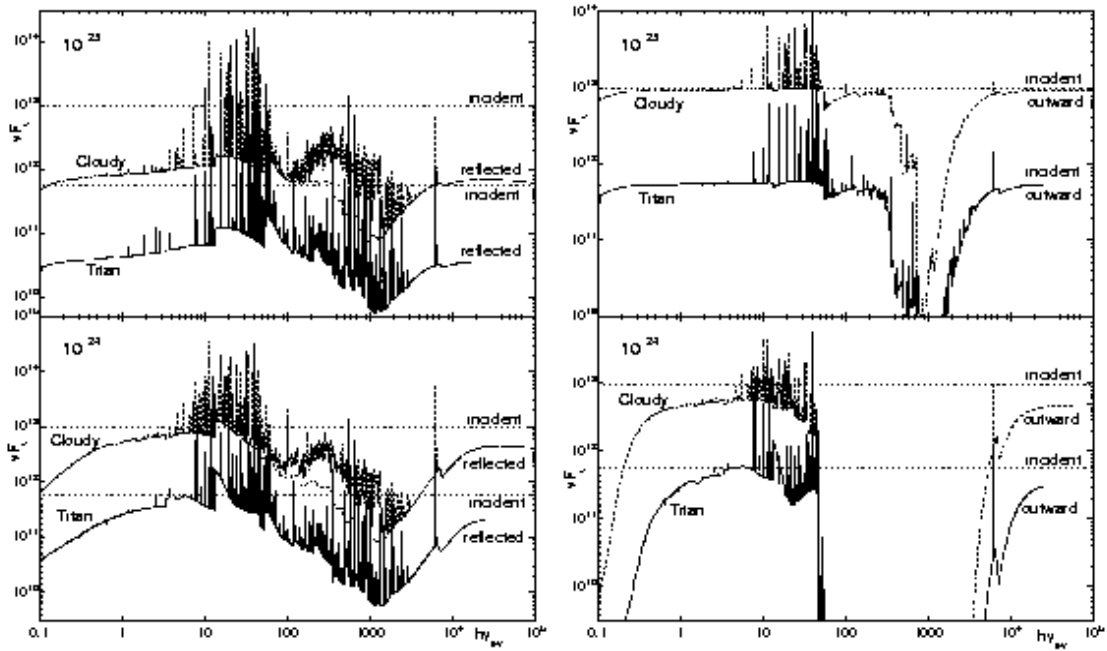,clip=,width=16.0cm}
\caption{Reflected and outward spectrum, for $\xi=100$ erg cm s$^{-1}$ and 
column densities equal to 10$^{23}$ and 10$^{24}$ cm$^{-2}$. The 
results obtained with {\sc{cloudy}} are shown for comparison in dashed 
lines. The incident flux of {\sc{cloudy}} is multiplied by a factor 
$\sqrt{3}$ to account for the semi-isotropy of the intensity
in {\sc{titan}}, and the emitted fluxes are  multiplied by an 
additional factor 10 for a better visibility. The spectra are 
displayed with a line resolution of 300. }
\label{fig-spectres-c23-24}
\end{figure*}

Fig. \ref{fig-spectres-c23-24} displays the reflected and outward spectra, 
for column densities 10$^{23}$ and 10$^{24}$ cm$^{-2}$.
We see that the overall shape of the continuum is very similar with 
{\sc{titan}} and with {\sc{cloudy}}. However, some differences in the detailed 
features appear for the largest column density. For instance the 
Lyman edge is in emission  in the reflected spectrum with both 
codes, but it is larger with {\sc{titan}} (presumably because the 
temperature is higher in the em-ission layers). In the outward spectrum 
it is present as a very weak absorption with {\sc{cloudy}}, while it is in 
emission with {\sc{titan}}. 

The line spectra are obviously also different. There are much more lines 
with {\sc{cloudy}}, but on the other hand the fewer lines of {\sc{titan}} 
are more intense. This is expected since subordinate lines are not 
taken into account exce-pt for hydrogen-like ions in {\sc{titan}}. As 
mentioned before, some lines appear in absorption with {\sc{titan}}, 
both in the reflected and in the outward spectrum. 
 
\subsection{Hot medium optically thick to Compton scattering: the 
UV-soft X-ray emitting medium in AGN}

{\sc{titan}} is specially designed for Compton thick photoionized media 
like those commonly assumed to emit the {UV/soft} X-ray continuum 
in AGN and to produce  through Compton reflection the FeK $\sim$7 KeV 
line and the 30 KeV hump observed in many Seyfert 1 galaxies. 

Although the exact nature of this medium is not known (an 
irradiated accretion disc, as proposed by Ross \& Fabian 1993, Zycki 
et al. 1994, to quote only the first papers on the subject, or a 
clumpy Compton thick medium, cf. Collin-Souffrin et al. 1996), its  
characteristics and physical state are comparable: a shell of gas 
with a column density of at least 10$^{25}$ cm$^{-2}$, a 
temperature of 10$^5$-10$^6$ K due to radiative heating, and a 
density spanning a range from 10$^{12}$ to 10$^{15}$ cm$^{-3}$. The 
corresponding value of $\xi$ is typically 300 to 3000 erg cm s$^{-1}$. 
Our reference model is therefore representative of this medium. In the case 
of an irradiated accretion disk, it has been mimicked as a sh-ell of 
constant density irradiated from above by a power law continuum, 
and from below by a black body radiation (actually it is not the best 
way to take into account the presence of the underlying viscously 
heated disk, cf. Rozanska et al. 2000).

%figure                                            18
\begin{figure}
\psfig{figure=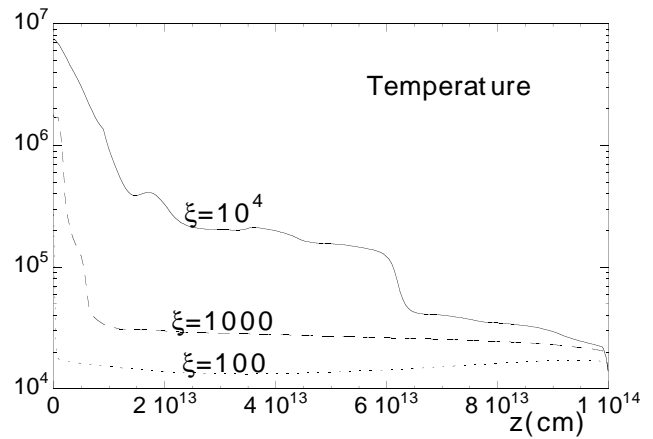,width=8.5cm}
\caption{Temperature in K versus $z$ for different values of $\xi$ in erg
cm s$^{-1}$, the other parameters being the same as in the reference model.}
\label{fig-Gtempc26x2-3-4}
 \end{figure}

%\begin{figure*}
%%\psfig{figure=ionisation-26-x2-x3.ps,width=16.0cm}
%\caption{}
%\label{fig-ionisation-26-x2-x3}
%\end{figure*}

Let us consider  first a slab illuminated on one side by a 
semi-isotropic radiation. Fig. \ref{fig-Gtempc26x2-3-4} displays $T$ 
versus $z$, for different values of $\xi$. The other parameters are 
the same as in the reference model. It is interesting to notice that 
the quasi constant temperature regime (corresponding to LTE) is 
reached for increasing values of $z$ when $\xi$ increases. For 
instance it is reached for $z\sim10^{13}$ cm, corresponding to 
$\tau_{\rm T}\sim 6$ for $\xi=1000$ erg cm s$^{-1}$. It means that any 
computation aiming at giving the reflected spectrum of such a model should 
{\it solve correctly the thermal and ionization equilibrium until these 
deep layers}. 

%figure                                            19
\begin{figure}
%\begin{figure*}
\psfig{figure=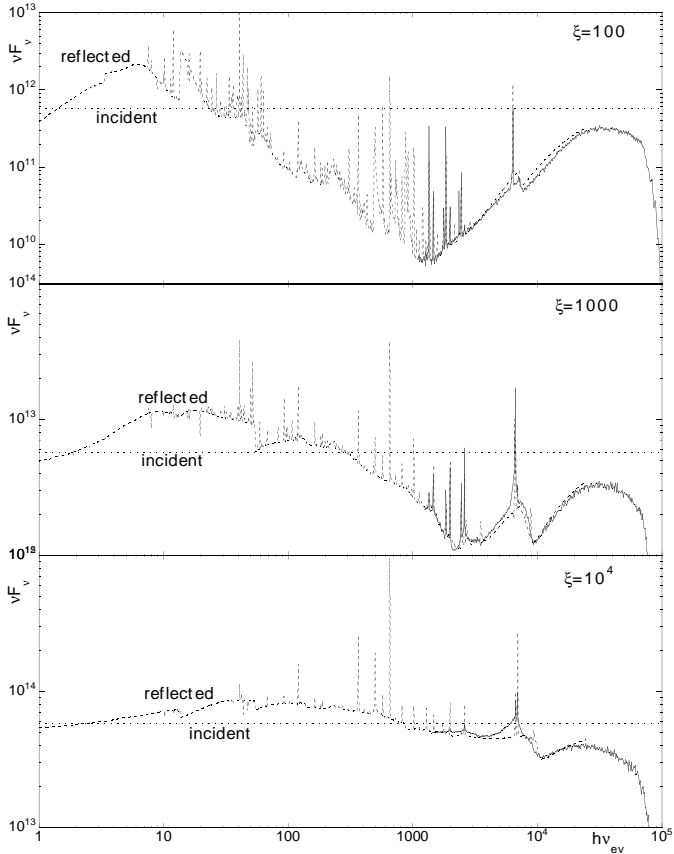,width=9.0cm}
\caption{Reflected spectrum for different values of $\xi$ in erg cm s$^{-1}$, 
the other parameters being the same as in the reference model; full lines: 
results of {\sc{noar}}; dashed lines: results of {\sc{titan}}. The spectra are 
displayed with a resolution of 100.}
\label{fig-Gtitnoarx2-3-4}
% \end{figure*}
 \end{figure}

Fig. \ref{fig-Gtitnoarx2-3-4} shows the reflected spectrum for the 
same models. Note that the resolution adopted here (100) does not 
correspond to a region where the lines are broadened by large 
velocity field, either turbulent or organized (rotation). We use this 
high resolution only to show more clearly the details of the spectrum.

We see that the overall shape of the spectrum computed with {\sc{titan}} 
and {\sc{noar}} is very similar in the range 1-20 KeV. However the 
detailed spectral features are different, particularly for the higher 
values of $\xi$. For instance the complex feature near 7 KeV (which 
is made of a mixture of several Iron edges and of Iron lines dominated by low 
ionization stages for $\xi=100$ erg cm s$^{-1}$, and by FeXXV and FeXXVI for 
$\xi=10^4)$ erg cm s$^{-1}$ is strongly modified by Compton 
{scattering}. For $\xi=10^4$ erg cm s$^{-1}$ the line has a large red wing 
due to direct Compton scattering, and a weak blue wing due to inverse Compton 
scattering. The absorption edge is also completely erased. 
The effect is smaller for $\xi=10^3$ erg cm s$^{-1}$, and almost absent 
for $\xi=100$ erg cm s$^{-1}$, as 
discussed above. Similar results have been obtained in many papers 
dealing with the X-ray spectrum of Seyfert galaxies (cf. for instance 
Ross et al. 1999).

%figure                                         20
\begin{figure}
\psfig{figure=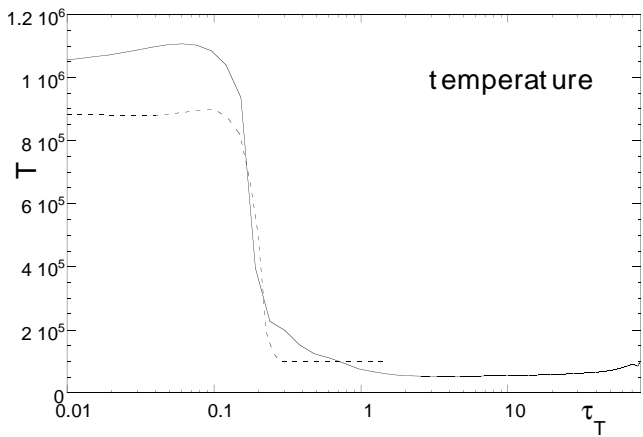,width=8.5cm}
\caption{Temperature in K versus $\tau_{\rm T}$ for the ``accretion disk" model
 (see text), computed by us (full line), and by Zycki et al. (dashed line).}
\label{fig-temp-Zyk}
 \end{figure}

%figure                                         21
\begin{figure}
\psfig{figure=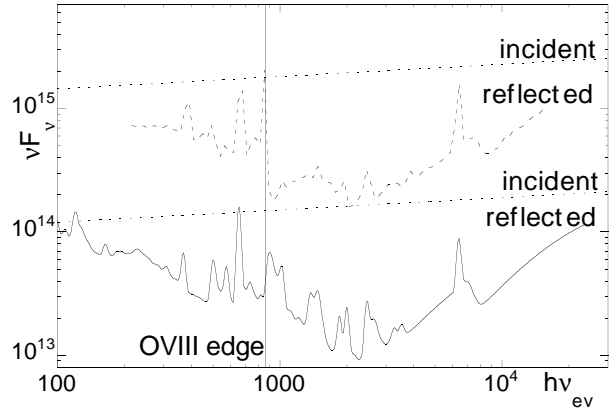,width=8.5cm}
\caption{Reflected spectrum for the ``accretion disk" model 
computed by us (full line), and by Zycki et al. (dashed 
line). Zycki et al. spectrum is shifted for clarity. The 
resolution is 20 for the {\sc{titan}} spectrum.}
\label{fig-spectre-Zyk}
 \end{figure}

Let us now compare the results obtained by us with {\sc{titan}}, and by 
Zycki et al. (1994), for a slab of constant density 10$^{14}$ 
cm$^{-3}$, illuminated by a power law $\nu^{-0.9}$ between 10 eV 
and 100 KeV on one side, and by a black body $T_{\rm BB}=10^5$K on 
the other side. We will compare the results for $\xi$(Zycki et al.)=300, 
which corresponds to $\xi$({\sc{titan}})$\sim$ 200 erg cm s$^{-1}$, owing to 
our semi-isotropic illumination, and to a slightly different definition 
of $\xi$. 

 Like us, Zycki et al. used a Monte-Carlo method  to take into
account Compton scattering. The thermal and ionization state of the 
gas is calculated apart with the code {\sc{xstar}}, using  like 
{\sc{cloudy}} the escape probability approximation.  
Fig. \ref{fig-temp-Zyk} gives $T$ versus  $\tau_{\rm T}$ obtained by us 
and by Zicky et al. The same result is observed as with {\sc{cloudy}}: our 
temperature is larger near the surface of the slab. It 
reaches $T_{\rm BB}$ at $\tau_{\rm T}$=0.7. For a larger value of 
$\xi$, the connection with the underlying black 
body would be reached in even deeper layers, as discussed 
above. For $\tau_{\rm T}\ge 0.7$ the temperature becomes smaller than 
$T_{\rm BB}$, and joins $T_{\rm BB}$ only at $\tau_{\rm T}\sim 20$. 
This behaviour is however dependent on the thickness of the slab, it is why 
this representation of the disk is not meaningful (Rozanska et al. 2000). 
On the contrary Zycki et al. impose the gas temperature to be equal 
to $min(T_{\rm BB},T_{\rm thermal\ balance})$.

Fig. \ref{fig-spectre-Zyk} displays the reflected spectrum, computed 
by us and by Zycki et al. The resolution is degraded to 20, to fit the 
energy bins of Zycki et al. The overall shape of the spectra are 
roughly similar. 
The major discrepancy is the OVIII edge, in emission with our computation 
and in absorption with Zycki et al.'s one. We think that it is due to the 
higher surface temperature and to the lower temperature in the deeper 
layers obtained in our computation.

\section{Conclusion}

We have shown that the coupling of {\sc{titan}} and {\sc{noar}} 
allows to compute the 
structure and the emission of hot Compton thick irradiated media in an 
unprecendented way. First it solves consistently both {\it the global and the 
local} energy balance, which is impossible in a thick medium with codes 
handling 
the line transfer with the escape probability approximation, as all present 
photoionization codes do. We have also shown the importance of the returning 
flux (which is neglected in photoionization codes) even for relatively low 
column densities. Second it takes into account in an exact way inverse and 
direct Compton scattering, both in the energy balance and in the computation of
the emitted spectrum. Finally, it allows to treat any geometry, open or closed.

Although the problem of the convergence process is not as drastic as in
stellar atmospheres, since we are able to get complete converged structure and
spectrum in a still reasonable computing time, a most urgent improvement of 
the code is to accelerate the convergence process through the use of the 
Accelerated Lambda Iteration method. This will not only allow to get the 
results in a much smaller time, but also to get convergence for the few 
lines which are still not converged after about 10$^3$ iterations.

Then the following improvements of {\sc{titan}} will be to take into account 
subordinate lines in solving a multi-level atom for all ions and to
bring up to date the atomic data. More
detailed line  spectra for some abundant ions will be obtained through 
coupling with the code  {\sc{iris}}. The L shells of Iron which are already 
taken into account for FeXVII to FeXXII will be implemented as well as a 
better representation of the Iron K  lines. A
few elements will be added to the already ten  existing ones. 

\begin{acknowledgements}

We are grateful to the referee, G. Ferland, for his comments which have led to
substantially clarify the paper, to M-C. Artru for a very careful reading 
of the manuscript, and to I. Hubeny for enlightning discussions.

\end{acknowledgements}

\section*{Appendix : Previous methods}

\subsection{Transfer of the continuum}

It is not always clear which approximations are used in the different codes.
Basically the computation of the diffuse radiation field seems to be 
similar in all codes, except in Ross \& Fabian (1993) and in subsequent 
works using this code, where it is
computed using the Kompaneets equation. {\sc{cloudy}} (Ferland 
1996), {\sc{xstar}} (Kallman \& Krolik, 1995 for the last version), 
{\sc{ion}} (Netzer 1990, 1993), use a modified version of the ``on the spot" 
approximation, which amounts to assuming a kind 
of escape probability for the diffuse continuum, and one 
stream approximation in the transfer equation.
 These approximations are correct only for a continuum optical 
thickness smaller than unity
(effective or total opacity), and when the properties of the cloud 
do not vary considerably between the point where a photon is emitted 
and the point where it is reabsorbed.
In particular one very important requirement for studying hot and 
thick media is to take into {account} the radiation returning from the 
backside of the slab, even when it is not illuminated. As the medium is
optically thick in a large frequency range, this radiation 
is intense and modifies the physical state of the whole slab, 
including the illuminated side(see Fig. \ref{fig-GTitanCloudyT-col}). 

\subsection{Line transfer: {\bf the escape probability approximation}}

All previous photoionization codes, except that of Collin-Souffrin 
and Dumont (S.) (1986), treat line transfer by the so-called local escape 
probability formalism,
which uses the probability that a line photon emitted at a given 
point can escape {\bf in a single flight}
from the cloud, assuming that the rest of the cloud is homogeneous 
and has the same properties as the emitting layer. It amounts to 
identifying the divergence flux of the statistical equilibrium equations 
with the escape probability
intervening in the line emerging flux. This approximation is valid
 only in the case of complete frequency
redistribution, absence of line interlocking, and {\bf if the medium is homogeneous}. Its use can have
severe consequences on the emission line spectrum and on the energy 
balance when these conditions are not fulfilled as it was often discussed,
see for example by Collin \& S.Dumont (1986) or Elitzur (1984).
We introduce here the divergence flux in the aim to show
that our results are completely different from those obtained by the
escape probability formalism.

To simplify the discussion, let us consider in the rest of this section
a simple two-level atom.  The level population balance writes:

\begin{eqnarray} 
n_u (A_{ul} \ +\ B_{ul}\int J_{\nu}{\psi}_{\nu}d\nu \ 
+\ n_{\rm e}  C_{ul})\\ 
\nonumber
=\ n_{l} (B_{lu}\int J_{\nu}{\phi}_{\nu} d\nu \ +\ n_{\rm e}C_{lu}) 
\label{eq-stat} \end{eqnarray}

\noindent where $\phi_{\nu}$ and $\psi_{\nu}$ are the absorption and 
emission line profiles, $A_{ul}$, $B_{ul}$ and $B_{lu}$, $C_{ul}$ and $C_{lu}$ 
are the usual radiative (Einstein) and collisional excitation and 
deexcitation coefficients, and $J_{\nu}$ is the angle averaged intensity. 
%(including the diffuse and the absorbed incident intensity).

Let us now define the {\it divergence flux} of the transition, $\rho _{ul}$:

\begin{equation} 
\rho _{ul}={\left\{{ n_u( A_{ul} + B_{ul}\int J_{\nu}{\psi}_{\nu}d\nu)
-  n_l (B_{lu}\int J_{\nu}{\phi}_{\nu} d\nu )}\right\} \over n_u A_{ul}}. 
\label{eq-div-flux} \end{equation}

With this definition Eq. \ref{eq-stat} becomes simply:

\begin{equation}  n_u\ (n_{\rm e} C_{ul}\ +\ {\rho}_{ul} A_{ul})\ 
=\ n_l\ n_{\rm  e}C_{lu}.
\label{eq-statbis} 
\end{equation}

%Now let us make the {\it approximation that the emission and the absorption
%profiles are equal}. Then the source function is constant on the 
%line profile and is equal to:
%\begin{equation} S\ =\ {{   n}_{   u}{   A}_{   ul } \over {   
%n}_{   l }{   B}_{
%  lu}}\ {1 \over 1\ -\ {{   n}_{   u}{   B}_{   ul } \over {   
%n}_{   l }{   B}_{lu}}} 
%\label{eq-source-fonct} \end{equation}
%\noindent and $\rho _{ul}$ can be written:
%\begin{equation} \rho {}_{   u\l }\ =\ 1\ -\ {\int_{}^{}{   
%J}_{\nu }{\phi }_{\nu}{ d}\nu  \over {   S}} 
%\label{eq-rho} \end{equation}
%This equation relates the emergent flux in a line to the
%divergence flux. The transfer equation, {\it assuming that the line is 
%the only contributor to the emissivity and absorption coefficient }, 
%integrated over frequencies and angles, writes:
%\begin{equation} {1 \over 4\pi }\ {dF \over dz}\ =\ \kappa \left(
%-\int_J_{\nu}{\phi}_{\nu} d{\nu}\ +\  S \right)\ 
%=\ \kappa \ {\rho}_{ul}S\ =\ {\rho}_{ul} \epsilon 
%\label{eq-dFdz} \end{equation}
%\noindent where $F$ is the local flux integrated over the line profile, 
%$\kappa$ is the mean line absorption coefficient, $\epsilon$ is the integrated
%emissivity $ n_u A_{ul}h\nu$, $dz$ is the differential geometrical 
%thickness. 

Let us assume {\it the line is the only contributor to radiation}.
The energy emitted locally in the line in a small volume of 
surface 1 cm$^{-2}$ and length $dz$ writes:

\begin{eqnarray}
\nonumber 
dF &=& h\nu\ [n_u (A_{ul} + B_{ul}\int J_{\nu}{\psi}_{\nu}d\nu) -
 n_l B_{lu}\int J_{\nu}{\phi}_{\nu} d\nu ]\ dz \\ 
&=& \rho _{ul} n_u A{ul}\ h\nu\ dz = \rho _{ul} 4\pi \epsilon dz
\label{eq-dFdz} \end{eqnarray}

where $\epsilon$ is the emissivity coefficient $ n_u A_{ul}h\nu /4\pi $.
Integrating this equation over depth, one gets the emerging line flux:

\begin{equation} F\ =\ 4\pi \int{\rho}_{ul}\ \varepsilon \ {dz} 
\label{eq-F} 
\end{equation}

According to this equation the divergence flux seems to be equal to the 
probability that, once emitted at a distance $z$ from the surface, a photon 
can escape from the medium, which is called the ``escape probability" $P_e$.

From this fact, if one makes {\it the approximation that the emission 
and the absorption profiles are equal}, one computes this escape 
probability by integrating the attenuation over the line profile:

\begin{equation} P_e\ \sim\ {1 \over 2}\ 
{\int_{0}^{1}\int_{}^{}{h\nu  \over 4\pi
}{   n}_{   u}\ {   A}_{ ul }\ {\phi }_{\nu }\ {   exp}(-
\tau {\phi }_{\nu }/\mu
)d\nu \ d\mu  \over \int_{}^{}{   h\nu  \over 4\pi }{   n}_{   
u}\ {   A}_{   ul}\ {\phi }_{\nu }\ {   d}\nu } 
\label{eq-esc-prob} 
\end{equation}

\noindent where $\mu$ = cos$\theta$, and it gives the usual expressions for 
complete redistribution in a Voigt profile:

\begin{equation}
 P_e(\tau_0) \sim max\left({1\over 1+2\tau_0\sqrt{\pi 
ln(\tau_0+1)}},{2\over 
3}\sqrt{{a\over\tau_0\sqrt{\pi}}}\right),
  \label{eq-esc-probter} 
  \end{equation}

\noindent where $\tau_0$ is the optical thickness at the line center
and $a$ is the usual damping constant. This expression can thus 
replace $\rho_{ul}$ in Eqs.  \ref{eq-statbis} and \ref{eq-F}.

%Note that Eq. \ref{eq-F}  can also be written, from Eq. \ref{eq-statbis},
%(still for the two-level case):
%\begin{equation} F\ = h\nu\
%\int_{}^{}[n_ln_eC_{lu}-n_un_eC_{ul}] {   dz} 
%\label{eq-F-HII} \end{equation}
%\noindent which is similar to that giving the emerging flux for a two level 
%atom, in dilute regions, such as HII regions, 
%planetary nebulae, or the Narrow Line Region of AGN. So  finally 
%$\rho_{ul}$ (i.e. $P_e$) enters only in the statistical  equilibrium equation.

With this method the frequency dependent transfer eq-uations are
successfully replaced by frequency integrated ones, and solved 
consistently with the statistical equat-ions. 

  Several estimates of the escape probability are used in the literature 
(see for example Kwan \& Krolik, 1981, Rees, Netzer \& Ferland, 1989, 
Kallman \& McCray, 1982, Ko \& Kallman, 1994), to account for partial 
redistribution, for line or continuum
interlocking, and for continuum absorption. For instance several 
ways to take into account the continuum
opacity (or the overlapping of two lines) have been proposed 
(Elitzur \& Netzer, 1984, Netzer, Elitzur \& Ferland, 1985). 
In the case of the Broad Line Region of AGN, which is optically thin for the 
continuum underlying the main lines, Collin-Souffrin et al. (1981), 
Avrett \& Loeser (1988), have discussed the
influence of different approximations for partial or complete 
redistribution, and how they compare to an exact
transfer treatment, and Collin-Souffrin
\& Dumont S. (1986) have shown that the escape probability 
approximation is roughly valid (it can however lead to a decrease 
by a factor as large as two of the ratio of a lower to a higher 
transition, such as H$\alpha$/H$\beta$ or L$\alpha$/H$\alpha$).

However these approximations are not
valid if ${\tau}_{\rm cont}$ is of the order of unity at the line 
frequency and if the line photons
are created in one place and absorbed in another place where the 
physical conditions are different. This happens for instance for EUV 
and soft X-ray lines created in the hot region and absorbed in ionizing 
He$^+$ ions in a cooler region.

If the local escape probability approximation breaks down,  it has a severe
consequence: {\it the emitted line spectrum and the thermal and 
the ionization balance are not correctly computed}. For instance 
the divergence flux takes frequently negative  values in our computations, 
even for intense lines, and even in the reflected spectrum. This is not 
allowed with escape probabilities. We 
illustrate  this discussion with some examples in Sect. 3.2.

\end{document}